\documentclass{article}

\usepackage[natbibapa]{apacite}
\usepackage{graphicx}
\usepackage{authblk}
\usepackage[section]{placeins}
\usepackage{longtable}
\usepackage{hyperref}

\usepackage[utf8]{inputenc} % allow utf-8 input
\usepackage[T1]{fontenc}    % use 8-bit T1 fonts
\usepackage{hyperref}       % hyperlinks
\usepackage{url}            % simple URL typesetting
\usepackage{booktabs}       % professional-quality tables
\usepackage{amsfonts}       % blackboard math symbols
\usepackage{nicefrac}       % compact symbols for 1/2, etc.
\usepackage{microtype}      % microtypography
\usepackage{xcolor}         % colors

\title{Effective Mitigations for Systemic Risks from General-Purpose AI}

\author[1,2]{Risto Uuk\thanks{Corresponding author: risto@futureoflife.org}}
\author[1]{Annemieke Brouwer}
\author[1]{Tim Schreier}
\author[3]{Noemi Dreksler}
\author[2]{Valeria Pulignano}
\author[4]{Rishi Bommasani\thanks{This work was initiated prior to, and is entirely unrelated to, the involvement of Rishi Bommasani in the EU AI Act Codes of Practice.}}
\affil[1]{Future of Life Institute}
\affil[2]{KU Leuven}
\affil[3]{Centre for the Governance of AI}
\affil[4]{Stanford University}

\date{14 November 2024}

\begin{document}

\maketitle

\begin{abstract}
The systemic risks posed by general-purpose AI models are a growing concern, yet the effectiveness of mitigations remains underexplored. Previous research has proposed frameworks for risk mitigation, but has left gaps in our understanding of the perceived effectiveness of measures for mitigating systemic risks. Our study addresses this gap by evaluating how experts perceive different mitigations that aim to reduce the systemic risks of general-purpose AI models. We surveyed 76 experts whose expertise spans AI safety; critical infrastructure; democratic processes; chemical, biological, radiological, and nuclear risks (CBRN); and discrimination and bias. Among 27 mitigations identified through a literature review, we find that a broad range of risk mitigation measures are perceived as effective in reducing various systemic risks and technically feasible by domain experts. In particular, three mitigation measures stand out: safety incident reports and security information sharing, third-party pre-deployment model audits, and pre-deployment risk assessments. These measures show both the highest expert agreement ratings (>60\%) across all four risk areas and are most frequently selected in experts' preferred combinations of measures (>40\%). The surveyed experts highlighted that external scrutiny, proactive evaluation and transparency are key principles for effective mitigation of systemic risks. We provide policy recommendations for implementing the most promising measures, incorporating the qualitative contributions from experts. These insights should inform regulatory frameworks and industry practices for mitigating the systemic risks associated with general-purpose AI.
\\ \\
\textbf{Keywords}: Systemic risk mitigation, general-purpose AI, expert survey, risk management. 

\end{abstract}

% \begin{figure}
%     \centering
%     \includegraphics[width=\linewidth]{Measures.png}
%     \caption{Caption}
%     \label{fig:enter-label}
% \end{figure}

\clearpage
\section*{Executive Summary}

The rapid development of general-purpose AI models poses systemic risks to society, ranging from risks to critical infrastructure and democratic processes to labour market impacts and environmental concerns. While risk mitigation measures have been proposed, their effectiveness and feasibility remains largely underexplored. This study addresses this gap by surveying 76 domain experts on their views of the effectiveness of 27 risk mitigation measures across four systemic risks. The central finding of the study is that experts perceive a wide range of mitigation measures to be both effective in reducing risks across four systemic risk areas and technically feasible.

\textbf{Methodology.} 76 domain experts across five key systemic risk areas – AI safety; critical infrastructure; democratic processes; chemical, biological, radiological, and nuclear risks (CBRN); and discrimination and bias – were surveyed. Experts evaluated the perceived effectiveness of 27 risk mitigation measures and could indicate if they thought the measure was not technically feasible. Domain experts were asked to assess the effectiveness of mitigation measures across four systemic risk categories specified in the AI Act: disruptions to critical sectors; negative effects on democratic processes; chemical, biological, radiological, and nuclear risks (CBRN); and harmful bias and discrimination. Experts were also asked to give rationales for their responses. The expert sample includes representation from academia (40.8\%), the non-profit sector (39.5\%), and industry (7.9\%), and was selected based on recognised contributions to the field, including influential publications, leadership roles in professional organisations, and participation in major AI conferences, while aiming for geographic and institutional diversity.

\textbf{Key findings of the study.}
\begin{itemize}
    \item A wide-range of measures were judged to be both effective and technically feasible. 
    \item An overwhelming majority of experts (91\% or above) thought all of the measures were technically feasible.
    \item Safety incident reporting and security information sharing emerged as the most effective measure across various risk domains by the experts (70-91\% agreement).
    \item Pre-deployment risk assessments and third-party pre-deployment audits were also among the highest ranked, emphasising the importance of external scrutiny (70-86\% and 62-87\% agreement, respectively).
\end{itemize}

\begin{figure}
    \centering
\includegraphics[width=\linewidth,height=0.8\textheight]{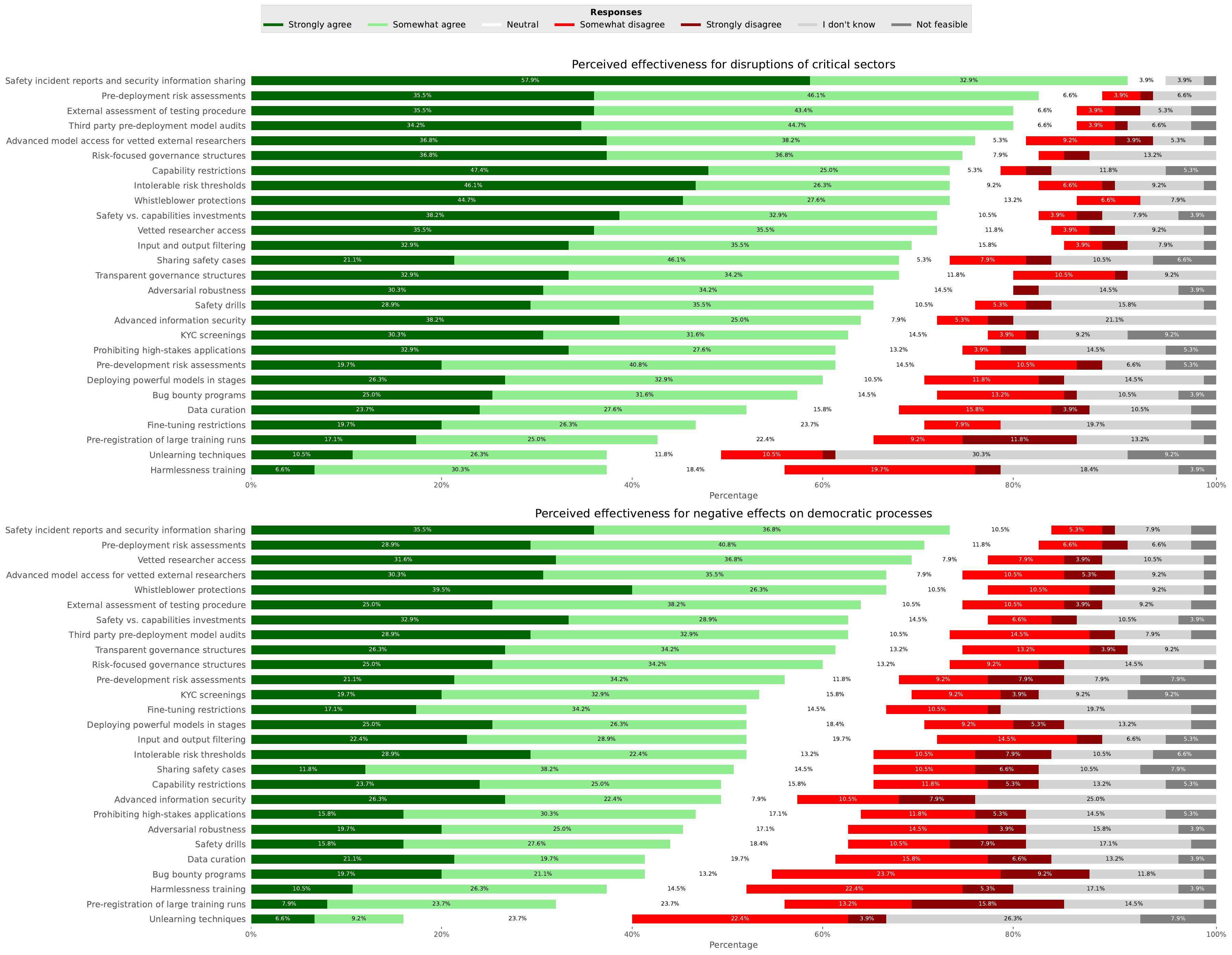}
    \caption{Expert agreement on effectiveness of different risk mitigation measures for general-purpose AI models across two systemic risks. Experts (n=76) were asked to what extent they agreed that the implementation of 27 risk mitigation measures by providers of large general-purpose AI models would effectively reduce four systemic risks from AI. Two are shown here: (1) Disruptions of critical sectors and (2) Negative effects on democratic processes. Experts were also able to indicate that a measure was not yet technically feasible.}
    \label{fig:risk1}
\end{figure}

\begin{figure}
    \centering
\includegraphics[width=\linewidth,height=0.8\textheight]{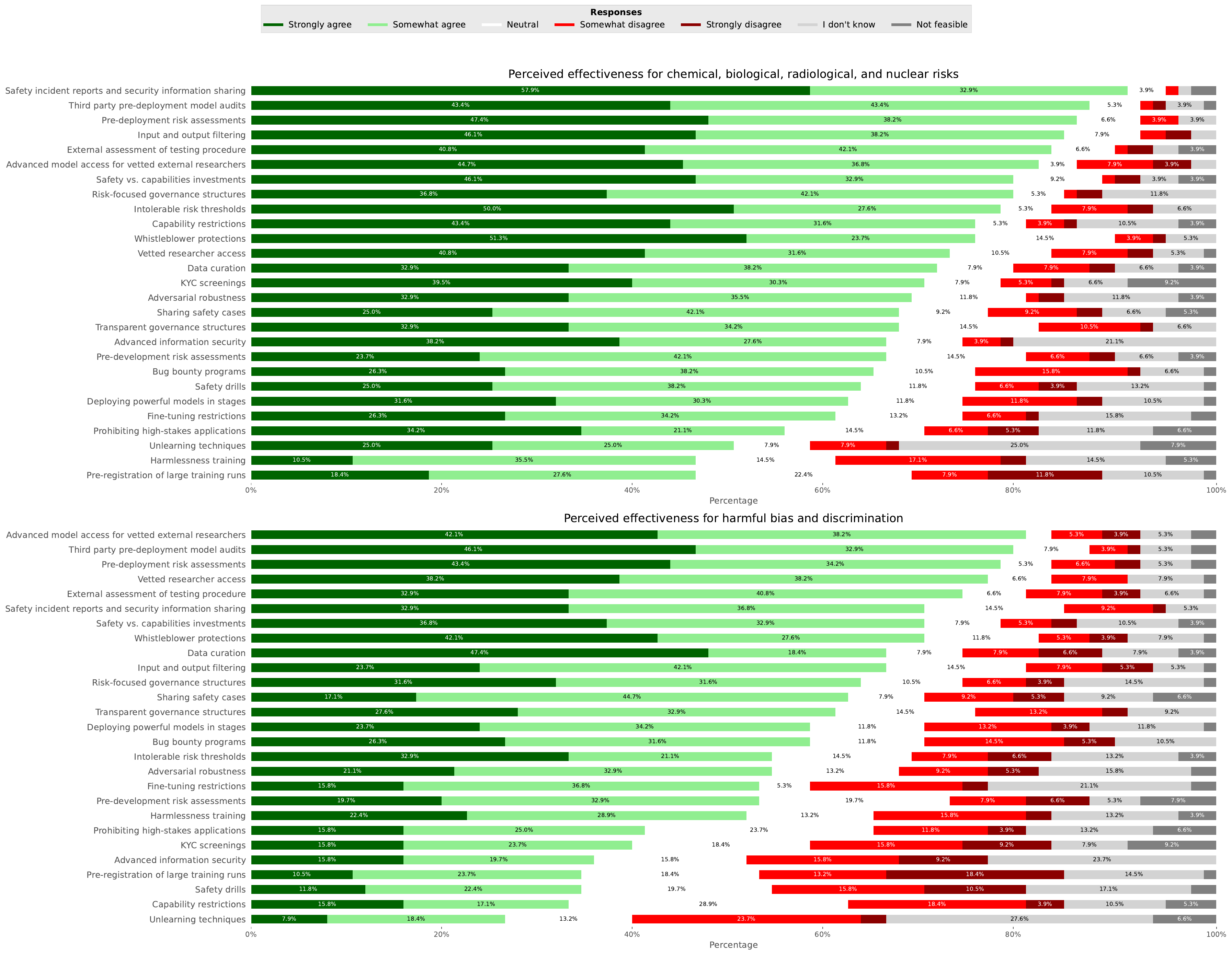}
    \caption{Expert agreement on effectiveness of different risk mitigation measures for general-purpose AI models across four systemic risks. Experts (n=76) were asked to what extent they agreed that the implementation of 27 risk mitigation measures by providers of large general-purpose AI models would effectively reduce four systemic risks. Two are shown here: (3) Chemical, biological, radiological, and nuclear risks (CBRN) and (4) Harmful bias and discrimination. Experts were also able to indicate that a measure was not yet technically feasible.}
    \label{fig:risk2}
\end{figure}

\textbf{Policy implications.} These findings have immediate policy relevance. In particular, the EU AI Act, which requires the development of compliance guidance for providers of general-purpose AI by May 2025. For one, the results suggest that experts think there are a wide-range of technically feasible and effective risk mitigation measures that could be implemented and legally mandated to reduce systemic risks. Our results suggest, in particular, that effective regulation should mandate robust reporting mechanisms, independent oversight, and multi-layered risk mitigation strategies.

The study identifies eight priority measures (details of the measures in \hyperref[sec:4.2]{Section 4.2}) for providers of general-purpose AI based on two criteria: how often experts selected them in their top-10 combinations and how strongly experts agreed with their effectiveness across all risk domains. Other measures also saw notable levels of agreement for specific risk areas, but these eight stand out across four risk categories:

\begin{enumerate}
    \item \textbf{Third-party pre-deployment model audits:} 62-87\% agreed that independent safety assessments of models before deployment, with auditors given appropriate access for testing, would effectively reduce systemic risk in each risk area. 98\% of experts thought the measure was technically feasible.
    \item     \textbf{Safety incident reporting and security information sharing:} 70-91\% agreed that disclosure of AI incidents, near-misses, and security threats to relevant stakeholders would effectively reduce systemic risk in each risk area. 98\% of experts thought the measure was technically feasible.
    \item \textbf{Whistleblower protections:} 66-75\% agreed that policies ensuring safe reporting of concerns without retaliation or restrictive agreements would effectively reduce systemic risk in each risk area. 99\% of experts thought the measure was technically feasible. 
    \item     \textbf{Pre-deployment risk assessments:} 70-86\% agreed that comprehensive assessment of potential misuse and dangerous capabilities before deployment would effectively reduce systemic risk in each risk area. 99\% of experts thought the measure was technically feasible. 
    \item \textbf{Risk-focused governance structures:} 59-79\% agreed that implementation of board risk committees, chief risk officers, multi-party authorisation requirements, ethics boards, and internal audit teams would effectively reduce systemic risk in each risk area. 99\% of experts thought the measure was technically feasible. 
    \item \textbf{Intolerable risk thresholds:} 51-78\% agreed that clear red lines for risk or model capabilities set by a third-party that trigger immediate development or deployment halt would effectively reduce systemic risk in each risk area. 97\% of experts thought the measure was technically feasible. 
    \item \textbf{Input and output filtering:} 51-84\% agreed that monitoring for dangerous inputs and outputs would effectively reduce systemic risk in each risk area. 98\% of experts thought the measure was technically feasible. 
    \item \textbf{External assessment of testing procedure:} 63-83\% agreed that third-party evaluation of how companies test for dangerous capabilities would effectively reduce systemic risk in each risk area. 97\% of experts thought the measure was technically feasible.
\end{enumerate}

\textbf{Key strengths and limitations of the study.} The strengths of the study are its mixed-methods approach, combining quantitative ratings with qualitative insights from expert input, and its qualified expert sample representing diverse expertise across five domains. The survey covered a broad range of risk mitigation measures that were drawn from an extensive literature review. The limitations of the study include potential response bias, assumption that measures would be legally required, well-executed, and overseen by competent regulators, and insufficient industry representation. Future research should examine larger samples, gather empirical evidence on the effectiveness of mitigation measures, and evaluate practical implementation challenges, such as resource constraints, regulatory capacity, and organisational culture variations.

This research provides actionable insights for policymakers and general-purpose AI providers for the development of regulatory frameworks and industry best practices for mitigating systemic risks from general-purpose AI.

\clearpage

\section{Introduction}

The rapid development of general-purpose AI models introduces risks with broad and far-reaching implications to society as a whole \citep{bommasani2022, clarke2021, fli2021, smuha2021}. Some research has detailed these implications, identifying threats such as discrimination and toxicity, information hazards, misinformation harms, malicious uses, human-computer interaction harms, and automation and environmental harms \citep{weidinger2021}. More recent work has addressed systemic risks ranging from labour market impacts to privacy risks and environmental concerns \citep{bengio2024}. These systemic risks illustrate the complex and diverse challenges that general-purpose AI models present, emphasising the need for effective and targeted mitigation strategies.

Recognising these challenges, the EU passed one of the first comprehensive AI regulations worldwide in 2024, the AI Act, which focuses on systemic risks, defining them as those stemming from the high-impact capabilities of general-purpose AI models, which can significantly affect the Union market due to their reach or reasonably foreseeable negative effects on public health, safety, public security, fundamental rights, or society as a whole (EU AI Act, Art. 3). These risks can proliferate widely across entire value chains, potentially manifesting as disruptions to critical sectors; negative effects on democratic processes; harmful bias and discrimination; and chemical, biological, radiological, and nuclear risks (CBRN) (EU AI Act, Rec. 110).

Despite growing recognition of these risks, current mitigation approaches remain understudied. Several mitigation frameworks have been proposed, including ethics guidelines for trustworthy AI \citep{eu2019}, algorithmic preparedness principles \citep{kolt2023}, safety processes \citep{ukgov2023}, technical risk mitigation approaches \citep{bengio2024}, model deployment guidance \citep{pai2023}, and risk management standards \citep{barrett2023}. However, these proposals lack systematic assessment of intervention effectiveness. The only study examining expert consensus on AI safety practices \citep{schuett2023a} focused broadly on AGI rather than specific systemic risks, and its purely quantitative methodology precluded deeper insights into expert reasoning and contextual factors, leaving gaps in our understanding of effective mitigation measures.

This knowledge gap has immediate policy implications. The European AI Office, as the main enforcer of the EU AI Act, is responsible for developing codes of practice to guide providers of general-purpose AI in compliance. These codes, due by May 2025, should be informed by research on effective risk mitigation measures, with a focus on systemic risks and best practices for ensuring safety, as well as the protection of public health and fundamental rights. Effectively addressing the systemic risks posed by general-purpose AI requires identifying and implementing the most effective risk mitigation measures. Understanding which mitigations best reduce risks will help policymakers and providers of general-purpose AI prioritise resources and design targeted and effective interventions. Such an approach involves a comprehensive evaluation of various risk mitigation strategies, assessing not only their potential effectiveness but also their feasibility.

In this paper, we aim to bridge these gaps by systematically evaluating the potential benefits of various risk mitigation measures, drawing from both literature review and expert opinion, to determine some of the most effective interventions for reducing systemic risks posed by general-purpose AI. Specifically, we survey the scientific literature on risk mitigations for providers of general-purpose AI and conduct a survey involving both AI safety experts and risk domain experts. By synthesising these findings, we aim to identify actionable, evidence-based measures to mitigate systemic risks from general-purpose AI.

\section{Methodology}
This section outlines the approach taken to identify and evaluate risk mitigation measures for general-purpose AI models. It details the survey design, participant selection criteria, sampling method, procedures used to collect and analyse data, and ethical considerations.
\subsection{Survey design}
\subsubsection{Identification of risk mitigation measures}
\label{sec:2.1.1}
This study began with a broad, narrative literature review to identify potential risk mitigation measures for providers of general-purpose AI. Drawing on our expertise and subjective judgement, we reviewed a variety of sources, including academic publications, industry reports, policy documents, and expert recommendations from the fields of AI safety, risk management, and technology governance. From this review, we curated a list of risk mitigation measures, based on our assessment of their potential to address systemic risks from general-purpose AI across various contexts. We established guiding criteria to determine which risk mitigation measures to include in the survey by asking of each whether it:

\begin{itemize}
    \item Addresses key drivers of systemic risk, either broadly or for specific risk types.
    \item Demonstrates a plausible connection to systemic risk reduction.
    \item Has been recommended by experts for similar risks in the past.
    \item Aligns with standard risk management best practices.
    \item Shows evidence of historical effectiveness in related contexts.
    \item Targets key bottlenecks or vulnerabilities in risk management processes.
    \item Strikes a balance between being too general and too specific, ensuring practical applicability across various scenarios.
    \item Represents practices that some companies have implemented but are not yet uniformly adopted across industries.
\end{itemize}

After the initial selection, the list of risk mitigation measures (see Appendix ~\ref{appendix:E} for the list) was further refined via a workshop at the Center for Human-Compatible AI Conference in California on 14 June 2024. Prior to the workshop, 15 attendees participated in a survey, providing ratings on an earlier selection of 37 risk mitigation measures based on their perceived effectiveness in reducing systemic risks from general-purpose AI. A full readout of the CHAI survey and workshop can be found in Appendix~\ref{appendix:A}. 

Table 1 presents a curated list of risk mitigation measures identified through an extensive literature review. After compiling this list, we designed an expert survey to evaluate the perceived effectiveness of each measure in mitigating systemic risks posed by general-purpose AI models.

\begin{longtable}{ |p{2.5cm}|p{4.5cm}|p{3cm}|  }
\caption{Risk mitigation measures, including their description and sources.} \\
\hline
\textbf{Risk mitigation measures} & \textbf{Description} & \textbf{Sources*} \\
\hline

Pre-development risk assessments
 & Comprehensive risk assessments based on forecasted capabilities \textit{before training }new models. Risk assessments would inform impactful development decisions. & \href{https://www-cdn.anthropic.com/1adf000c8f675958c2ee23805d91aaade1cd4613/responsible-scaling-policy.pdf}{Anthropic, 2023}, p. 13; \href{https://assets.publishing.service.gov.uk/media/653aabbd80884d000df71bdc/emerging-processes-frontier-ai-safety.pdf}{UK Government, 2023}, p. 10; \href{https://doi.org/10.48550/arXiv.2305.15324}{Shevlane et al., 2023}, p. 9; \href{https://arxiv.org/pdf/2303.08774}{OpenAI, 2023b}, p. 59. \\
\hline
Pre-deployment risk assessments
 & Comprehensive risk assessments before deployment that would assess reasonably foreseeable misuse and include dangerous capability evaluations that incorporate post-training enhancements and collaborations with domain experts. Risk assessments would inform deployment decisions. & \href{https://cdn.governance.ai/AGI_Safety_Governance_Practices_GovAIReport.pdf}{Schuett et al., 2023a}, p. 12, 18; \href{https://partnershiponai.org/wp-content/uploads/1923/10/PAI-Model-Deployment-Guidance.pdf}{Partnership on AI, 2023}, p. 3; \href{https://doi.org/10.48550/arXiv.2404.02675}{Kolt et al., 2024}, p. 4; \href{https://doi.org/10.48550/arXiv.2305.15324}{Shevlane et al., 2023}, p. 9; \href{https://cltc.berkeley.edu/publication/ai-risk-management-standards-profile}{Barrett et al., 2023}, p. 21; \href{https://assets.publishing.service.gov.uk/media/653aabbd80884d000df71bdc/emerging-processes-frontier-ai-safety.pdf}{UK Government, 2023}, p. 17; \href{https://arxiv.org/pdf/2312.07413}{Davidson et al., 2023}, p. 22. \\
\hline

Third party pre-deployment model audits
 & External pre-deployment assessment to provide a judgement — or input to a judgement — on the safety of a model. Auditors, which could be governments or independent third parties, would receive access to a fine-tuning API for testing, or further appropriate technical means. & \href{https://www.whitehouse.gov/wp-content/uploads/2023/07/Ensuring-Safe-Secure-and-Trustworthy-AI.pdf}{The White House, 2023}, p. 3; \href{https://partnershiponai.org/wp-content/uploads/1923/10/PAI-Model-Deployment-Guidance.pdf}{Partnership on AI, 2023}, p. 3; \href{https://cdn.governance.ai/AGI_Safety_Governance_Practices_GovAIReport.pdf}{Schuett et al., 2023a}, p. 18; \href{https://assets.publishing.service.gov.uk/media/653aabbd80884d000df71bdc/emerging-processes-frontier-ai-safety.pdf}{UK Government, 2023}, p. 17. \\
\hline

External assessment of testing procedure
 & Bringing in external AI evaluation firms before deployment to assess and red-team the company's execution of dangerous capabilities evaluations. & \href{https://papers.ssrn.com/sol3/papers.cfm?abstract_id=4467502}{Skoric, 2023}. \\
\hline

Vetted researcher access
 & Giving good faith, public interest evaluation researchers access to black-box research APIs that provide technical and legal safe harbours to limit barriers imposed by usage policy enforcement, logging, and stringent terms of service. & \href{https://doi.org/10.48550/arXiv.2305.15324}{Shevlane et al., 2023}, p. 6, \href{https://assets.publishing.service.gov.uk/media/653aabbd80884d000df71bdc/emerging-processes-frontier-ai-safety.pdf}{UK Government, 2023}, p. 36; \href{https://cdn.governance.ai/AGI_Safety_Governance_Practices_GovAIReport.pdf}{Schuett et al., 2023}, p. 19; \href{https://arxiv.org/pdf/2403.04893}{Longpre et al., 2024, p. 7}. \\
\hline

Advanced model access for vetted external researchers
 & Examples of advanced access rights could include any of the following: increased control over sampling, access to fine-tuning functionality, the ability to inspect and modify model internals, access to training data, or additional features like stable model versions. & \href{https://assets.publishing.service.gov.uk/media/653aabbd80884d000df71bdc/emerging-processes-frontier-ai-safety.pdf}{UK Government, 2023}, p. 18, \href{https://cdn.governance.ai/Structured_Access_for_Third-Party_Research.pdf}{Bucknall \& Trager, 2023}, p. 12. \\
\hline

Data curation
 & Careful data curation prior to all development stages (including fine-tuning) to filter out high-risk content and ensure the training data is sufficiently high-quality. & \href{https://arxiv.org/pdf/2312.11805}{Gemini Team et al., 2024}, p. 28; \href{https://cltc.berkeley.edu/publication/ai-risk-management-standards-profile}{Barrett et al., 2023}, p. 29, 56; \href{https://assets.publishing.service.gov.uk/media/653aabbd80884d000df71bdc/emerging-processes-frontier-ai-safety.pdf}{UK Government, 2023}, p. 43. \\
\hline

Harmlessness training
 & State-of-the-art reinforcement learning and fine-tuning techniques, such as Reinforcement Learning from Human Feedback (RLHF) or Direct Preference Optimization (DPO), to ensure models do not engage in unsafe behaviour. & \href{https://assets.publishing.service.gov.uk/media/653aabbd80884d000df71bdc/emerging-processes-frontier-ai-safety.pdf}{UK Government, 2023}, p. 39; \href{https://www-cdn.anthropic.com/de8ba9b01c9ab7cbabf5c33b80b7bbc618857627/Model_Card_Claude_3.pdf}{Anthropic, 2024a}, p. 10; \href{https://arxiv.org/pdf/2312.11805}{Gemini Team et al., 2024}, p. 29; \href{https://arxiv.org/pdf/2303.08774}{OpenAI, 2023b}, p. 61. \\
\hline

Adversarial robustness
 & State-of-the-art methods such as adversarial training to make models robust to adversarial attacks (e.g., jailbreaking). & \href{https://cltc.berkeley.edu/publication/ai-risk-management-standards-profile}{Barrett et al., 2023}, p. 18; \href{https://arxiv.org/pdf/2312.11805}{Gemini Team et al., 2024}, p. 36. \\
\hline

Unlearning techniques
 & Removing specific harmful capabilities (e.g., pathogen design) from models using unlearning techniques. & \href{https://arxiv.org/pdf/2403.03218}{Li et al., 2024}; \href{https://arxiv.org/abs/2408.00761}{Tamirisa et al., 2024} \\
\hline

Deploying powerful models in stages
 & Starting with a small number of applications and fewer users, and gradually scaling up API-access as rigorous monitoring increases confidence in the model’s safety. An API-mediated staged release would also be required before open-sourcing a model. & \href{https://cdn.governance.ai/AGI_Safety_Governance_Practices_GovAIReport.pdf}{Schuett et al., 2023a}, p. 19; \href{https://cltc.berkeley.edu/publication/ai-risk-management-standards-profile}{Barrett et al., 2023}, p. 59, 76; \href{https://arxiv.org/pdf/2302.04844}{Solaiman, 2023}. \\
\hline

Fine-tuning restrictions
 & Restricting or closely monitoring fine-tuning access to models to ensure safeguards remain intact. & \href{https://doi.org/10.48550/arXiv.2310.00328}{O'Brien et al., 2023}, p. 12. \\
\hline

Capability restrictions
 & Restricting risky capabilities of deployed models, such as advanced autonomy (e.g., self-assigning new sub-goals, executing long-horizon tasks) or tool use functionalities (e.g., function calls, web browsing). & \href{https://doi.org/10.48550/arXiv.2310.00328}{O'Brien et al., 2023}, p. 12, 13. \\
\hline

KYC screenings
 & Know-your-customer (KYC) screenings before granting access to models with very high misuse potential or to users producing large amounts of output. & \href{https://cdn.governance.ai/AGI_Safety_Governance_Practices_GovAIReport.pdf}{Schuett et al., 2023a}, p. 19; \href{https://assets.publishing.service.gov.uk/media/653aabbd80884d000df71bdc/emerging-processes-frontier-ai-safety.pdf}{UK Government, 2023}, p. 39; \href{https://openai.com/index/disrupting-malicious-uses-of-ai-by-state-affiliated-threat-actors/}{OpenAI, 2024a}. \\
\hline

Prohibiting high-stakes applications
 & Enforcing use policies that prohibit high-stakes applications. Requires Know-Your-Customer procedures. & \href{https://doi.org/10.48550/arXiv.2310.00328}{O'Brien et al., 2023}, p. 10, 12; \href{https://cdn.governance.ai/AGI_Safety_Governance_Practices_GovAIReport.pdf}{Schuett et al., 2023a}, p. 19, \href{https://assets.publishing.service.gov.uk/media/653aabbd80884d000df71bdc/emerging-processes-frontier-ai-safety.pdf}{UK Government, 2023}, p. 39; \href{https://openai.com/index/disrupting-malicious-uses-of-ai-by-state-affiliated-threat-actors/}{OpenAI, 2024b}. \\
\hline

Input and output filtering
 & Monitoring for dangerous outputs (e.g., code that appears to be malware or viral genome sequences) and inputs that violate acceptable use policies to ensure models do not engage in harmful behaviour. & \href{https://doi.org/10.48550/arXiv.2310.00328}{O'Brien et al., 2023}, p. 12, 24; \href{https://assets.publishing.service.gov.uk/media/653aabbd80884d000df71bdc/emerging-processes-frontier-ai-safety.pdf}{UK Government, 2023}, p. 38. \\
\hline

Bug bounty programs
 & Clear and user-friendly bug bounty programs that acknowledge and reward individuals for reporting model vulnerabilities and dangerous capabilities. & \href{https://cltc.berkeley.edu/publication/ai-risk-management-standards-profile}{Barrett et al., 2023}, p. 27, 48; \href{https://cdn.governance.ai/AGI_Safety_Governance_Practices_GovAIReport.pdf}{Schuett et al., 2023a}, p. 18; \href{https://assets.publishing.service.gov.uk/media/653aabbd80884d000df71bdc/emerging-processes-frontier-ai-safety.pdf}{UK Government, 2023}, p. 4, 29; \href{https://www.whitehouse.gov/wp-content/uploads/2023/07/Ensuring-Safe-Secure-and-Trustworthy-AI.pdf}{The White House, 2023}, p. 3. \\
\hline

Safety drills
 & Regularly practising the implementation of an emergency response plan to stress test the organisation’s ability to respond to reasonably foreseeable, fast-moving emergency scenarios. & \href{https://cdn.governance.ai/AGI_Safety_Governance_Practices_GovAIReport.pdf}{Schuett et al., 2023a}, p. 19; \href{https://cltc.berkeley.edu/publication/ai-risk-management-standards-profile}{Barrett et al., 2023}, p. 60. \\
\hline

Intolerable risk thresholds
 & Assessing and monitoring AI models with regard to red-line risk or capability thresholds set by a third-party, such as a standardisation organisation or regulator. Companies would further need to make technical, legal, and organisational preparations to halt development and deployment immediately when a breach occurs. & \href{https://cltc.berkeley.edu/publication/ai-risk-management-standards-profile}{Barrett et al., 2023}, p. 32, 54; \href{https://assets.publishing.service.gov.uk/media/653aabbd80884d000df71bdc/emerging-processes-frontier-ai-safety.pdf}{UK Government, 2023}, p. 11-12; \href{https://doi.org/10.48550/arXiv.2310.00328}{O'Brien et al., 2023}, p. 20, 24; \href{https://arxiv.org/pdf/2406.14713}{Koessler et al., 2024}. \\
\hline

Risk-focused governance structures
 & Companies adopt practices typical of high-reliability organisations (HROs), including board risk committees, chief risk officers, multi-party authorisation requirements, ethics boards for reviewing development and deployment decisions, and internal audit teams that report directly to the board, tasked with auditing risk management practices. & \href{https://cdn.governance.ai/AGI_Safety_Governance_Practices_GovAIReport.pdf}{Schuett et al., 2023a}, p. 19; \href{https://arxiv.org/pdf/2304.07249}{Schuett et al., 2023b}; \href{https://cltc.berkeley.edu/publication/ai-risk-management-standards-profile}{Barrett et al., 2023}, p. 22, 40; \href{https://partnershiponai.org/wp-content/uploads/1923/10/PAI-Model-Deployment-Guidance.pdf}{Partnership on AI, 2023}, p. 3; \href{https://assets.publishing.service.gov.uk/media/653aabbd80884d000df71bdc/emerging-processes-frontier-ai-safety.pdf}{UK Government, 2023}, p. 14; \href{https://www.whitehouse.gov/wp-content/uploads/2023/07/Ensuring-Safe-Secure-and-Trustworthy-AI.pdf}{The White House, 2023}, p. 2; \href{https://doi.org/10.48550/arXiv.2404.02675}{Kolt et al., 2024}, p. 4; \href{https://doi.org/10.48550/arXiv.2305.15324}{Shevlane et al., 2023}, p. 6. \\
\hline

Whistleblower protections
 & Refraining from restrictive non-disparagement agreements and instantiating comprehensive whistleblower protection policies that clearly outline relevant reporting processes, protection mechanisms, and non-retaliation assurances. & \href{https://cdn.governance.ai/AGI_Safety_Governance_Practices_GovAIReport.pdf}{Schuett et al., 2023a}, p. 21; \href{https://cltc.berkeley.edu/publication/ai-risk-management-standards-profile}{Barrett et al., 2023}, p. 24. \\
\hline

Safety vs. capabilities investments
 & A significant fraction of employees and computational resources are dedicated to enhancing model safety rather than advancing its capabilities. & \href{https://cdn.governance.ai/AGI_Safety_Governance_Practices_GovAIReport.pdf}{Schuett et al., 2023a}, p. 11, 18. \\
\hline

Safety incident reports and security information sharing
 & Disclosing reports about AI incidents, such as concrete harms and near misses as well as cyber threat intelligence and security incident reports with appropriate stakeholders such as select governments. & \href{https://assets.publishing.service.gov.uk/media/653aabbd80884d000df71bdc/emerging-processes-frontier-ai-safety.pdf}{UK Government, 2023}, p. 20; \href{https://partnershiponai.org/wp-content/uploads/1923/10/PAI-Model-Deployment-Guidance.pdf}{Partnership on AI, 2023}, p. 4; \href{https://doi.org/10.48550/arXiv.2404.02675}{Kolt et al., 2024}, p. 4, 7; \href{https://doi.org/10.48550/arXiv.2310.00328}{O'Brien et al., 2023}, p. 24; \href{https://cdn.governance.ai/AGI_Safety_Governance_Practices_GovAIReport.pdf}{Schuett et al., 2023a}, p. 18, 19; \href{https://doi.org/10.48550/arXiv.2305.15324}{Shevlane et al., 2023}, p. 9; \href{https://www.whitehouse.gov/wp-content/uploads/2023/07/Ensuring-Safe-Secure-and-Trustworthy-AI.pdf}{The White House, 2023}, p. 2. \\
\hline

Sharing safety cases
 & Disclosing to a regulator how high-stakes decisions regarding model development and deployment are made. & \href{https://arxiv.org/pdf/2403.10462}{Clymer et al., 2024}, p. 14; \href{https://www.aisi.gov.uk/work/safety-cases-at-aisi}{Irving, 2024}; \href{https://www-cdn.anthropic.com/1adf000c8f675958c2ee23805d91aaade1cd4613/responsible-scaling-policy.pdf}{Anthropic, 2023}, p. 15; \href{https://storage.googleapis.com/deepmind-media/DeepMind.com/Blog/introducing-the-frontier-safety-framework/fsf-technical-report.pdf}{Google DeepMind, 2024}, p. 4. \\
\hline

Transparent governance structures
 & Disclosing to a regulator how high-stakes decisions regarding model development and deployment are made. & \href{https://cdn.governance.ai/AGI_Safety_Governance_Practices_GovAIReport.pdf}{Schuett et al., 2023a}, p. 19. \\
\hline

Pre-registration of large training runs
 & Registering upcoming training runs above a certain size with an appropriate state actor. Such reports could include descriptions of architecture, training compute, data collection and curation, training objectives and techniques, and planned risk management procedures. & \href{https://cdn.governance.ai/AGI_Safety_Governance_Practices_GovAIReport.pdf}{Schuett et al., 2023a}, p. 20. \\
\hline

Advanced information security
 & Implementing advanced cybersecurity measures and insider threat safeguards to protect proprietary and unreleased model weights. & \href{https://assets.publishing.service.gov.uk/media/653aabbd80884d000df71bdc/emerging-processes-frontier-ai-safety.pdf}{UK Government, 2023}, p. 24; \href{https://cltc.berkeley.edu/publication/ai-risk-management-standards-profile}{Barrett et al., 2023}, p. 47; \href{https://www.whitehouse.gov/wp-content/uploads/2023/07/Ensuring-Safe-Secure-and-Trustworthy-AI.pdf}{The White House, 2023}, p. 3; \href{https://cdn.governance.ai/AGI_Safety_Governance_Practices_GovAIReport.pdf}{Schuett et al., 2023a}, p. 12, 18; \href{https://www-cdn.anthropic.com/de8ba9b01c9ab7cbabf5c33b80b7bbc618857627/Model_Card_Claude_3.pdf}{Anthropic, 2024a}, p. 26; \href{https://doi.org/10.48550/arXiv.2305.15324}{Shevlane et al., 2023}, p. 9. \\
\hline
\end{longtable}

\textit{*Note: Some sources were used for conceptual inspiration, and the descriptions provided may differ from the original texts. These sources were not shown to participants during the survey.}

After the initial selection, we further refined the list of measures through consultation with AI safety experts at the Center for Human-Compatible AI Conference in California \footnote{In an earlier phase of this study, we hosted a workshop at the Center for Human-Compatible AI Conference in California on June 14, 2024. Prior to the workshop, 15 participants completed a survey in which they rated 37 risk mitigation measures based on their effectiveness in reducing systemic risks from general-purpose AI. Using the insights from this preliminary survey, we refined our list of risk mitigation measures. 
}. 

\subsubsection{Survey questions}
The survey was designed to quantitatively assess the perceived effectiveness of each risk mitigation measure while also gathering qualitative insights into the reasoning behind expert evaluations. It also aimed to identify potential gaps or additional measures not captured in the initial literature review. Participants were introduced to the survey with an overview of systemic risks as defined by the EU AI Act and asked to evaluate the selection of risk mitigation measures listed above in  \hyperref[sec:2.1.1]{Section 2.1.1}. 

We divided the general-purpose AI-related systemic risks mentioned in the AI Act into the following four categories:

\begin{enumerate}
    \item \textbf{Disruptions of critical sectors}, such as digital infrastructure or other critical infrastructure, including via cyberattacks
    \item \textbf{Negative effects on democratic processes}, such as civil discourse and electoral procedures
    \item \textbf{Chemical, biological, radiological, and nuclear risks}, and other serious risks to public health and safety
    \item \textbf{Harmful bias and discrimination} with risks to communities or societies
\end{enumerate}

Our survey focused on the large general-purpose AI models which the EU AI Act classifies as having systemic risks. Under the EU AI Act, general-purpose AI models are considered to have systemic risks when the cumulative amount of computation used for its training measured in floating point operations exceeds $10^{25}$ FLOP. At the time of the survey, prominent examples likely included GPT-4o, Gemini Ultra 1.5, Claude 3.5, and Llama 3.1 405B.

The survey included both closed-ended and open-ended questions to gather a comprehensive understanding of each measure's perceived effectiveness. Closed-ended questions required participants to rate the effectiveness of each measure in reducing specific systemic risks, using a 5-point Likert scale ranging from "Strongly disagree" to "Strongly agree," along with options for "I don’t know" and "Not feasible." Additionally, participants were asked to select a combination of ten measures they estimated would be most effective at reducing systemic risks from general-purpose AI. The open-ended questions allowed participants to elaborate on their ratings and selected combinations, as well as to propose additional mitigation strategies. Participants were instructed to assume that each measure was legally required, well-executed, and overseen by a competent regulator. This approach ensured that their responses focused on the inherent value of the measures, rather than potential challenges related to implementation or enforcement. A full list of the survey questions, as well as the information presented to participants, is provided in Appendix~\ref{appendix:B}.

\subsection{Sampling method}
\subsubsection{Participant selection}
The target population for this survey comprised 303 experts in the fields of AI safety; critical infrastructure and AI; democratic processes and AI; chemical, biological, radiological, and nuclear risks (CBRN) and AI; and discrimination and bias in AI. The target population included academics, industry professionals, policymakers, and representatives from non-governmental organisations (NGOs) with demonstrated expertise in these areas. A non-probability sampling approach was used, with a purposive sampling method because our aim was to identify the most relevant experts to answer the research question.

We used purposive sampling. Experts were selected based on their recognised contributions to the field, including influential publications, leadership roles in professional organisations, and participation in major AI conferences. We aimed for representation from diverse geographical regions and institutional backgrounds to ensure a wide range of perspectives.

The list of potential participants was compiled from a range of sources, including the following:

\begin{itemize}
    \item Authors of key academic papers identified through Google Scholar
    \item Papers that have seemed to be influential to the authors in academic and policy discussions
    \item Recognised experts recommended by peers and colleagues within the field
    \item Individuals known through professional networks
\end{itemize}

\subsubsection{Survey distribution}
Experts were contacted via email and social media platforms. The communication included a clear explanation of the survey’s purpose, estimated time commitment, and assurances of confidentiality.

\FloatBarrier

\subsubsection{Sample characteristics}
The goal was to achieve a diverse sample that captured a wide range of expertise and perspectives within the field. While a specific sample size was not predetermined due to the exploratory nature of the research, efforts were made to include a broad representation across geographic regions, professional sectors (academia, industry, NGOs, think tanks), and demographic characteristics (gender, location).  \hyperref[tab:sample-comparison]{Table 2} provides an overview of the sample characteristics.

\begin{table}[h!]
\caption{Target and realised sample characteristics}
\label{tab:sample-comparison}
\centering
\begin{tabular}{l rr}
\toprule
\textbf{Category} & \textbf{Target sample} & \textbf{Realised sample} \\
\midrule
Sample size & 303 participants & 76 participants \\
\addlinespace
Gender & & \\
\quad Male & 53.8\% & 53.9\% \\
\quad Female & 46.2\% & 36.8\% \\
\quad Non-binary & -- & 3.9\% \\
\quad Prefer not to say & -- & 5.3\% \\
\addlinespace
Professional field & & \\
\quad Academia & 37.6\% & 40.8\% \\
\quad Non-profit & 34.0\% & 39.5\% \\
\quad Industry & 17.2\% & 7.9\% \\
\quad Other & 11.2\% & 11.8\% \\
\addlinespace
Area of expertise & & \\
\quad AI safety & 35.6\% & 38.2\% \\
\quad Critical infrastructure and AI & 12.2\% & 10.5\% \\
\quad Democratic processes and AI & 21.1\% & 17.1\% \\
\quad CBRN risks and AI & 14.2\% & 15.8\% \\
\quad Discrimination and bias in AI & 16.8\% & 18.4\% \\
\addlinespace
Seniority level & & \\
\quad Junior level & 11.9\% & 18.4\% \\
\quad Mid level & 55.0\% & 56.6\% \\
\quad Senior level & 33.1\% & 25.0\% \\
\addlinespace
Publications & & \\
\quad With relevant publications & 94.4\% & 92.1\% \\
\addlinespace
Current residence & & \\
\quad EU & 21.1\% & 25.0\% \\
\quad US & 54.1\% & 40.8\% \\
\quad UK & 15.5\% & 22.4\% \\
\quad Other & 9.2\% & 11.8\% \\
\bottomrule
\end{tabular}
\end{table}

\subsubsection{Representativeness and limitations}
Although the sampling approach limits the ability to generalise findings to the entire expert population, steps were taken to ensure diversity and inclusivity within the sample. The use of peer recommendations and broad sample selection strategies aimed to mitigate potential biases inherent in non-probability sampling.

The survey achieved a response rate of 25\% (76 out of 303 targeted participants), which may indicate a potential response bias in addition to any limitations of the selection of the sample. Comparing the demographic characteristics between target and realised samples reveals some patterns.

Regarding representation from professional sectors, academia is slightly overrepresented (40.8\% vs 37.6\% in target). The non-profit sector is also overrepresented (39.5\% vs 34.0\%). Industry is notably underrepresented (7.9\% vs 17.2\%). This indicates stronger participation from academic and non-profit sectors, and these sectors may show greater concern about AI risks compared to industry practitioners.

With regard to area of expertise, AI safety experts have slightly more representation than intended (38.2\% vs 35.6\%). Chemical, biological, radiological, and nuclear risks (CBRN) experts show higher participation as well (15.8\% vs 14.2\%). Experts in democratic processes show lower participation (17.1\% vs 21.1\%), but discrimination and bias experts slightly higher participation (16.8\% vs 18.4\%). This pattern does not offer a simple explanation, but it is possible that there was stronger engagement from those focused on technical risk mitigation measures, which were a strong component of the survey.

We can offer some possible explanations without confidence in any of them because of a small sample size. The lower industry participation might mean we are missing important practical implementation perspectives, including with regard to feasibility of risk mitigation measures. The high academic and non-profit representation could skew responses toward more theoretical or cautionary viewpoints. Finally, the strong response from AI safety experts might indicate overrepresentation of those most concerned about AI risks. 

\subsection{Data analysis}
In the survey, experts provided quantitative input for each topic first, followed by the option to elaborate with qualitative rationales. This structure allowed us to gather both objective assessments and contextual insights for a comprehensive understanding of each risk mitigation measure.

\subsubsection{Quantitative data analysis}
The quantitative data, collected from Likert scale and ranking responses, were analysed to assess the perceived effectiveness of each risk mitigation measure across the four systemic risk categories.

The data includes ratings from all experts across all four risk areas. This data was used to generate stacked bar charts visualising expert agreement on the effectiveness of each risk mitigation measure (see \hyperref[fig:risk1]{Figure 3} and \hyperref[fig:risk2]{Figure 4}). The ranking responses were also analysed to determine how often each measure was selected as part of the experts' top-ten combinations. The results were visualised using a bar plot, which displayed the selected measures broken down by expert group affiliations. Experts were given the option to rate measures as "not feasible" if they believed the scientific or technical understanding of the intervention was insufficient for robust implementation. The data was processed to produce a feasibility graph, highlighting the proportion of experts who deemed certain measures currently infeasible.

\subsubsection{Qualitative data analysis}
Qualitative data from the open-ended responses were analysed thematically to identify key insights and rationales behind expert opinions. The analysis involved reviewing around 120 pages of qualitative input, with a combination of manual reading and assistance from a general-purpose AI model, Claude Sonnet 3.5, to help identify patterns and develop concise summaries. This analysis helped identify specific factors influencing expert perspectives on feasibility and effectiveness, as well as uncover any gaps or additional mitigation measures not captured during the initial review. It is important to note that there are potential concerns about using an AI model to analyse the qualitative data; we only used it for initial pattern identification and making summaries more concise. The qualitative data and summaries underwent thorough human review, but mistakes could have been still made due to issues like hallucinations with these models. 

\subsection{Ethical considerations}
The survey was conducted in accordance with ethical guidelines for research involving human subjects. Informed consent was obtained from all participants, who were assured of the confidentiality and anonymity of their responses.

\section{Results}
Our survey of domain experts revealed varying levels of perceived effectiveness for different systemic risk mitigation measures. The results are summarised in three key visualisations: a stacked bar chart displaying the agreement levels for each measure by risk area (\hyperref[fig:risk1]{Figure 3} and \hyperref[fig:risk2]{Figure 4}), and a stacked bar chart displaying how often each risk mitigation measure was selected as part of experts' top ten choices, broken down by expert group affiliations (\hyperref[fig:combinations]{Figure 5}). In addition to the quantitative analysis, we also reflect extensively on the qualitative insights provided by experts.

\subsection{Expert agreement on effectiveness of measures}
\subsubsection{Quantitative analysis of expert opinion}
\hyperref[fig:risk1]{Figure 3} and \hyperref[fig:risk2]{Figure 4} present expert opinions on the effectiveness of various risk mitigation measures for general-purpose AI models across four risk areas. The results reveal a diverse range of perceived effectiveness among the proposed measures. The insights likely reflect the specific challenges and priorities within risk areas.

\textit{Safety incident reports and security information sharing} emerges as the most agreed to be effective measure, with strong agreement across all risks. Some measures show more variation in perceived effectiveness across different risks. For instance, \textit{advanced information security }and \textit{prohibiting high-stakes applications} receive mixed responses. It is worth noting that in the case of some risk areas like disruptions of critical sectors most measures get support from over half of the experts, suggesting that experts see potential value in a wide array of risk mitigation measures.

\subsubsection{Qualitative analysis of expert opinion}
Summaries of the qualitative expert insights for each measure are presented in Appendix~\ref{appendix:C}. The feedback provided by the experts covers a wide range of considerations for each measure, including:

\begin{itemize}
    \item Potential effectiveness across different risk areas
    \item Implementation challenges
    \item Limitations and potential drawbacks
    \item Comparisons with practices in other industries
    \item Suggestions for improvements or complementary measures (see also \hyperref[sec:other-measures]{Section 3.4} for an overview and Appendix~\ref{appendix:D} for the complete list)
\end{itemize}

\begin{figure}
    \centering
\includegraphics[width=\linewidth,height=0.8\textheight]{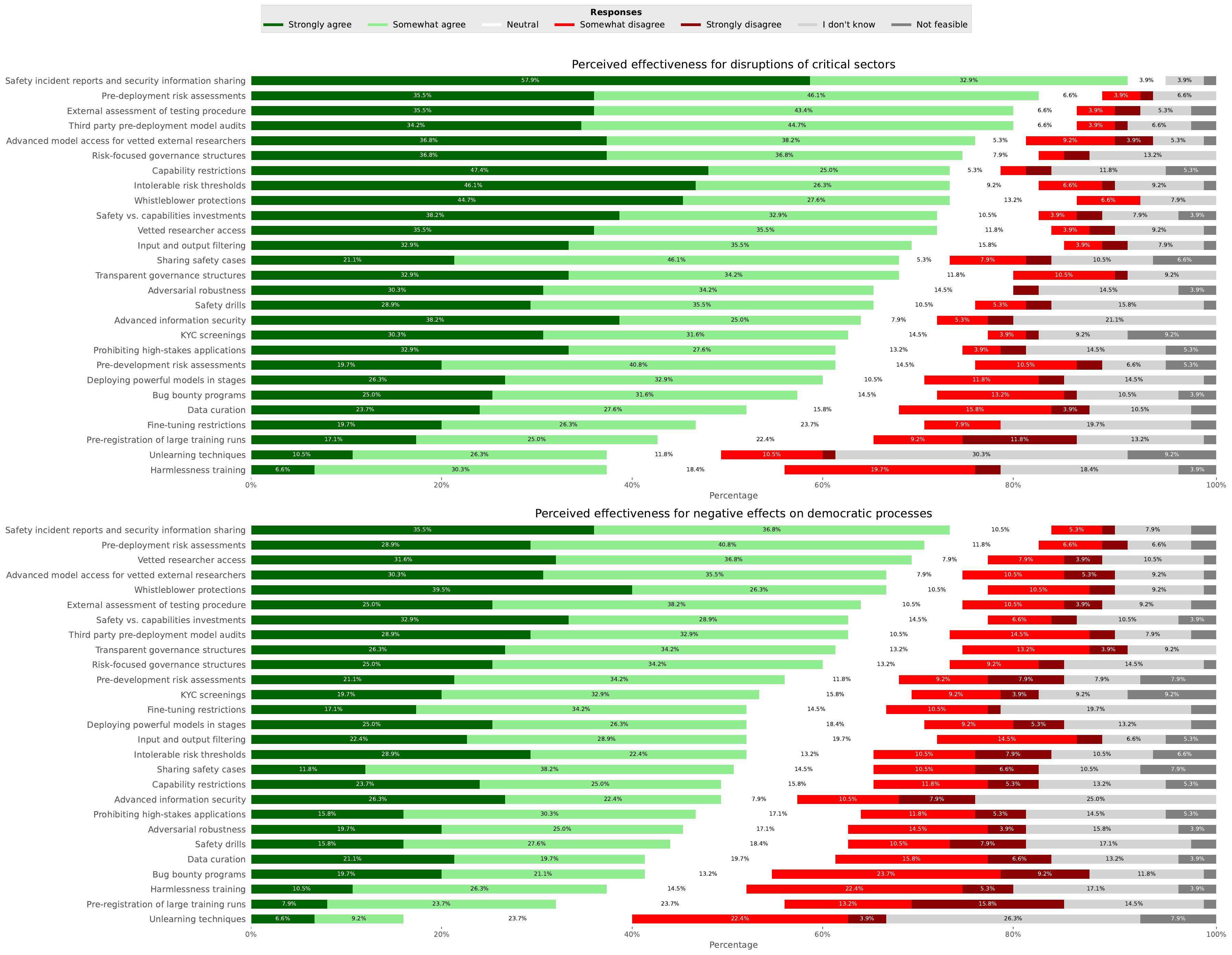}
    \caption{Expert agreement on effectiveness of different risk mitigation measures for general-purpose AI models across two systemic risks. Experts (n=76) were asked to what extent they agreed that the implementation of 27 risk mitigation measures by providers of large general-purpose AI models would effectively reduce four systemic risks from AI. Two are shown here: (1) Disruptions of critical sectors and (2) Negative effects on democratic processes. Experts were also able to indicate that a measure was not yet technically feasible.}
    \label{fig:risk1}
\end{figure}

\begin{figure}
    \centering
\includegraphics[width=\linewidth,height=0.8\textheight]{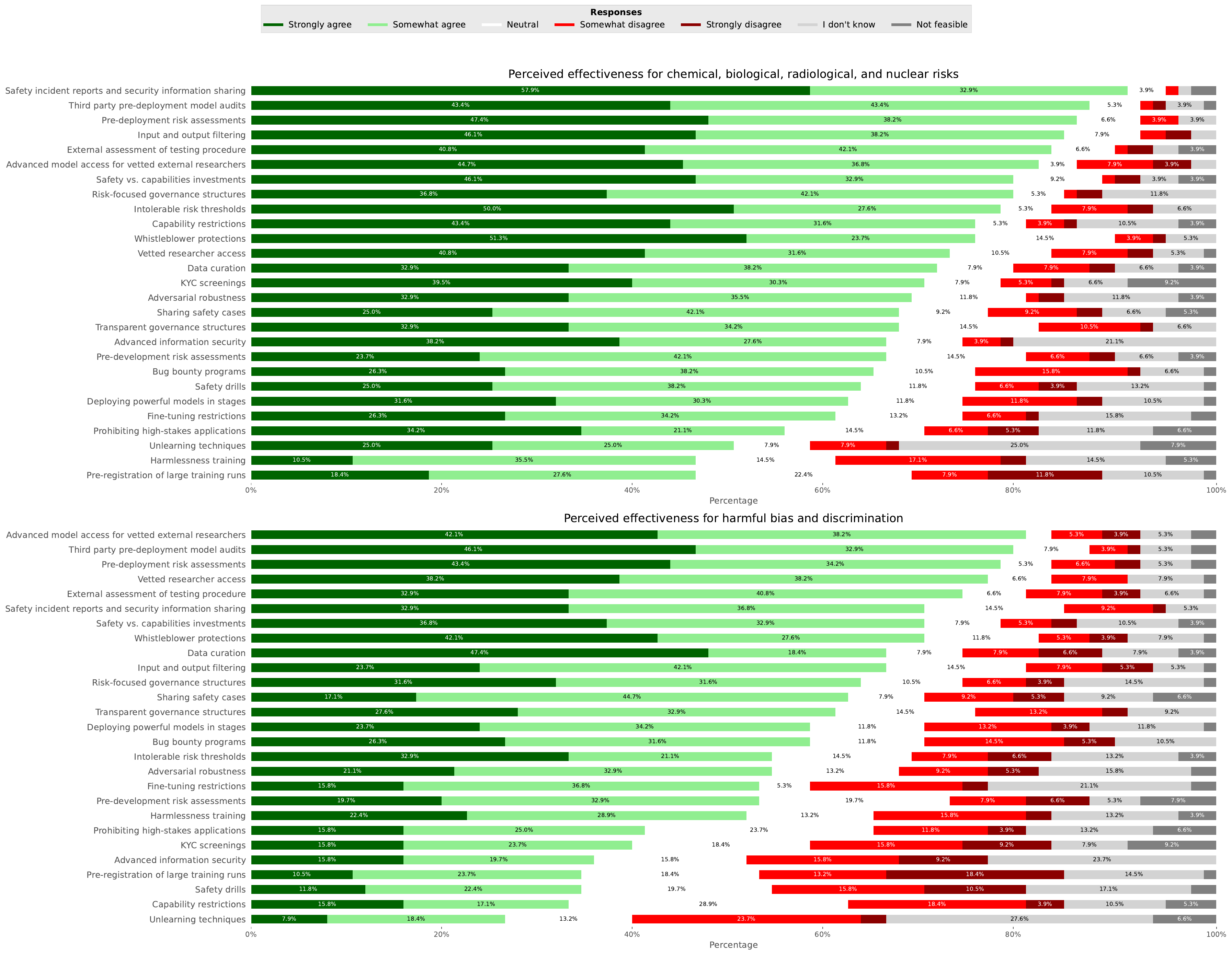}
    \caption{Expert agreement on effectiveness of different risk mitigation measures for general-purpose AI models across four systemic risks. Experts (n=76) were asked to what extent they agreed that the implementation of 27 risk mitigation measures by providers of large general-purpose AI models would effectively reduce four systemic risks. Two are shown here: (3) Chemical, biological, radiological, and nuclear risks (CBRN) and (4) Harmful bias and discrimination. Experts were also able to indicate that a measure was not yet technically feasible.}
    \label{fig:risk2}
\end{figure}

Below are concise summaries of recurring themes and notable insights from expert feedback on the three highest-scoring measures on both expert agreement ratings and selection in top 10 measures frequency, as identified in our quantitative analysis:

\begin{enumerate}
    \item \textbf{Safety incident reports and security information sharing}: Many experts viewed this measure favourably across various risk domains, drawing parallels to successful reporting systems in industries such as aviation. They highlighted its value in addressing security-related risks, including chemical, biological, radiological, and nuclear risks (CBRN); and disruptions to critical infrastructure. Several experts stressed the importance of timely reporting and the inclusion of "near-misses" to enhance proactive risk management. Additionally, clear processes and incentives for reporting were deemed crucial by some experts. However, one expert cautioned about the potential for information overload if not properly managed.
\item \textbf{Third-party pre-deployment model audits}: Experts emphasised the potential of third-party audits to provide unbiased perspectives and improve risk identification across different risk categories. Many mentioned the value of incorporating red-teaming exercises into these assessments to identify vulnerabilities more effectively. However, challenges were also noted, including the lack of standardised evaluation methods and the critical need for auditor independence and expertise. Some experts mentioned that audits alone may be insufficient to address complex, long-term, or systemic risks.
    \item \textbf{Pre-deployment risk assessments}: Experts generally agreed that pre-deployment risk assessments could be beneficial across all risk areas as a best practice. Many highlighted the importance of involving domain experts to ensure comprehensive and informed evaluations. However, several experts cautioned that the effectiveness of these assessments is highly dependent on who conducts them and their level of independence. Concerns were raised about potential biases in internal assessments, particularly due to conflicts of interest favouring deployment. Additionally, some experts noted the inherent limitations of these assessments, emphasising the challenges of predicting all potential risks, especially those that are unforeseen or rapidly evolving.
\end{enumerate}

\subsection{Expert opinion on combinations of risk mitigation measures}
\subsubsection{Quantitative analysis of preferred measures}
\begin{figure}
    \centering
    \includegraphics[width=\linewidth]{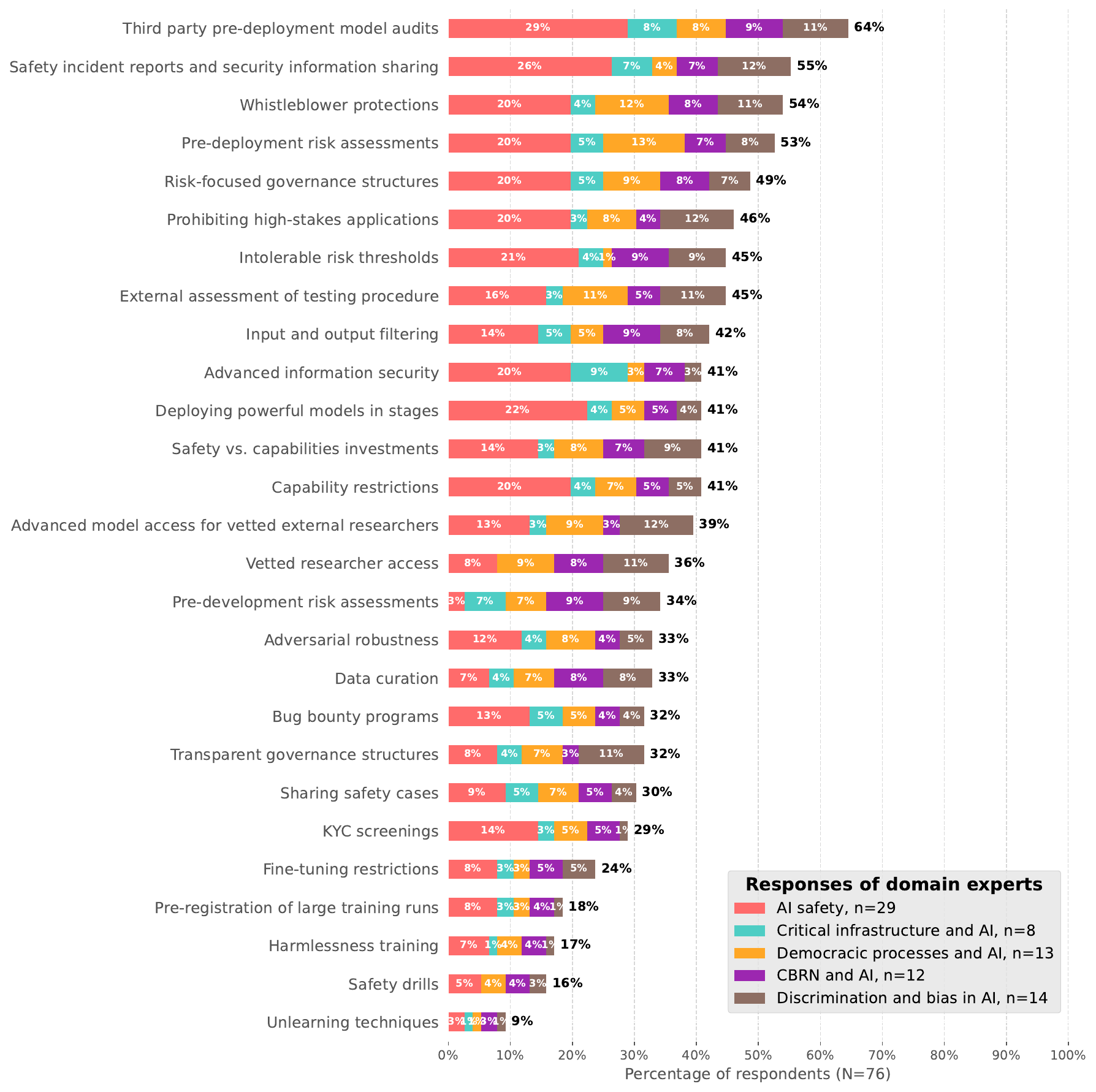}
    \caption{Most selected measures for reducing systemic risk effectively. Experts were asked what combination of ten measures they believed would be most effective at reducing the systemic risks from general-purpose AI. The Figure shows the percentage of the total experts (n=76) who included each measure in their ten measures (indicated at the end of each bar). The colour of the bar represents the expert group, though note that there were different numbers of experts in each group so that the label of each bar segment does not represent the percentage of the expert group who chose the measure.}
    \label{fig:combinations}
\end{figure}

We asked experts to identify the ten measures they believed to be most effective in reducing systemic risks from general-purpose AI. \hyperref[fig:combinations]{Figure 5} presents how often these measures were selected, broken down by domain expert group, and then summarised across all respondents. The results indicate that almost two-thirds of respondents chose third party pre-deployment model audits in their preferred list of measures, highlighting it as a favoured mitigation measure.

When comparing the combination rankings to individual rankings, some measures consistently appeared at the top of both lists, indicating expert consensus. These highly ranked measures include \textit{third party pre-deployment model audits, safety incident reports and security information sharing, whistleblower protections,} and p\textit{re-deployment risk assessments}. Their frequent selection underscores their perceived importance across different contexts. Some measures shifted in priority when selected in combination rather than individually. For instance, \textit{prohibiting high-stakes applications} ranked higher in the combination list, suggesting that experts find it more effective when paired with other measures. 

Individual and combined rankings showed some differences in the prioritisation of measures. Some measures ranked higher when evaluated individually than in combination with others. Measures like \textit{advanced model access for vetted external researchers} and \textit{vetted researcher access} had a higher priority in the individual ranking. \textit{Fine-tuning restrictions} and \textit{transparent governance structures} were occasionally also ranked higher in the individual list compared to the combinational rankings, possibly indicating that while valuable individually, their perceived contribution may diminish when combined with other mitigation efforts. However, given the complexity of the survey, these ranking differences might reflect response inconsistencies rather than substantive preferences.

\subsubsection{Qualitative analysis of expert rationales}
Experts provided diverse rationales for their selection of the top ten risk mitigation measures. Many emphasised the practicality of measures that could feasibly be implemented and subjected to external scrutiny, favouring those that involved governance, oversight, and independent evaluation. There was significant support for prioritising information sharing, transparency, and security practices, with several experts highlighting parallels to best practices in information security, where the focus is on limiting the damage of potential attacks. Some experts underscored the importance of iterative, ex ante approaches that prevent risks from escalating, while others expressed a preference for measures that slow the introduction of powerful AI capabilities to society. Several participants noted that measures aimed at structured transparency and risk assessment, such as third-party audits and whistleblower protections, were crucial for enabling external accountability and mitigating risks that companies might overlook due to internal biases.

Additionally, several experts expressed concerns about over-reliance on specific technical interventions, advocating instead for a broader mix of governance, transparency, and risk management strategies to address the complexity of systemic AI risks effectively. Overall, the qualitative feedback suggested that a well-rounded combination of technical, organisational, and regulatory measures – integrated with independent oversight – would be necessary for effectively managing systemic risks from general-purpose AI.

\subsection{Feasibility of risk mitigation measures}
\subsubsection{Quantitative analysis of feasibility}
We gave experts the option to rate measures as "not feasible" if they believed it is not feasible to robustly implement a certain measure because the scientific or technical understanding of the intervention is not mature enough. \hyperref[fig:feasibility]{Figure 6} shows the percentage of experts that rated measures as not feasible. The figure shows that all measures have at least 91\% or more of the experts assuming that it is currently sufficiently feasible to implement them. The measures that are considered infeasible by the highest number of experts are: \textit{KYC screenings} (9\%), \textit{unlearning techniques} (8\%) and \textit{sharing safety cases} (7\%). These measures also scored relatively low on both individual effectiveness and occurrence in preferred combinations of measures.

\subsubsection{Qualitative analysis of feasibility}
\begin{figure}
    \centering
    \includegraphics[width=\linewidth]{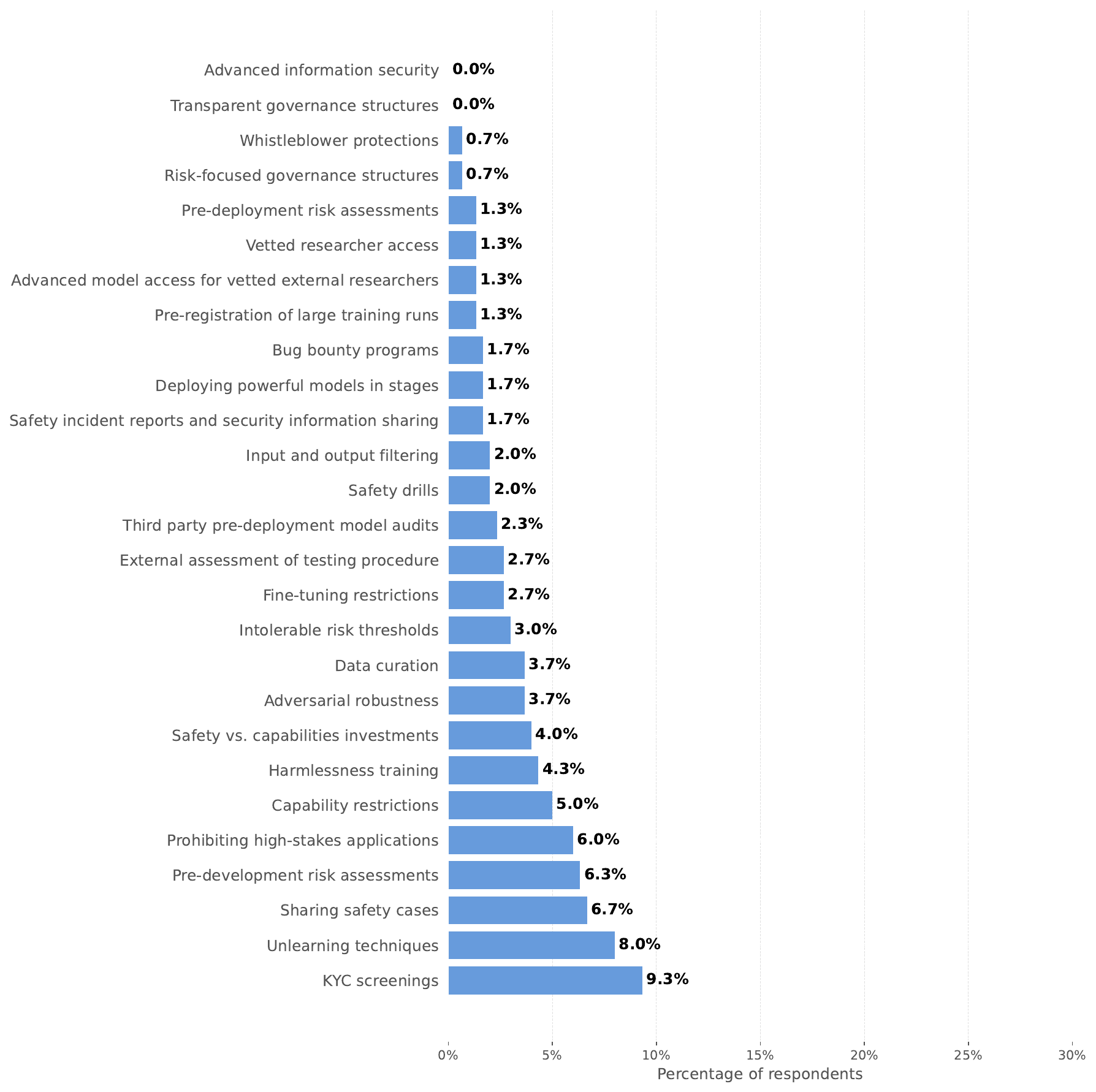}
    \caption{Proportion of experts who indicated a risk mitigation measure is not currently feasible with regards to the scientific or technical understanding of the intervention.}
    \label{fig:feasibility}
\end{figure}

Out of 76 experts, only 24 provided at least one comment on the feasibility of the risk mitigation measures. Therefore, these views are not fully representative of the entire expert group and may differ from the quantitative analysis presented in \hyperref[fig:feasibility]{Figure 6}.

Experts provided mixed views on the feasibility of various risk mitigation measures, often pointing to challenges related to the current state of scientific and technical knowledge. Some experts expressed scepticism regarding \textit{pre-development risk assessments}, citing limitations in forecasting capabilities and the difficulty of predicting specific downstream risks prior to training. The consensus was that these assessments, while promising, are not yet supported by sufficiently mature methods, particularly in areas like systemic risks from general-purpose AI models. For \textit{pre-deployment risk assessments}, several experts indicated that these could be feasible for certain risks, such as those involving chemical, biological, radiological, and nuclear (CBRN) risks, but they raised concerns about biases when assessments are conducted internally by companies. \textit{Third-party audits} were seen as valuable due to their external nature, although experts noted that they may face challenges in effectively identifying biases in the model or slowly evolving risks.

The feasibility of \textit{unlearning techniques} was questioned by experts, who viewed them as too nascent and lacking sufficient robustness. They highlighted the difficulty of applying unlearning to broad and complex issues, such as biases or impacts on democratic processes, due to the interlinked nature of knowledge within AI models. Concerns about the recoverability of unlearned capabilities and the limited applicability to open-source models further complicated the perceived feasibility. Measures like \textit{advanced model access for vetted researchers} and \textit{data curation} were also met with reservations. Experts emphasised that the technical understanding required to implement these interventions effectively is still in development. For \textit{data curation}, filtering training data to address specific risks, such as bias or chemical, biological, radiological, and nuclear (CBRN) risks, was seen as labour-intensive and potentially detrimental to overall model capabilities.

\textit{Safety drills} were generally regarded as effective for acute risks but were seen as less applicable to slow-developing threats like biases or democratic impacts. Experts noted difficulties in simulating realistic scenarios for these types of risks, raising doubts about the overall utility of such drills for certain systemic risks. Ultimately, experts highlighted the importance of balancing these measures with comprehensive oversight and iterative improvements. They also noted the need for further research to refine and enhance the feasibility of various risk mitigation techniques.

\subsection{Other suggested measures}
\label{sec:other-measures}
Experts suggested several additional measures beyond those included in the survey (for a complete overview, see Appendix~\ref{appendix:D}). One suggestion was the implementation of coordinated red-teaming to thoroughly assess model vulnerabilities, akin to penetration testing for security infrastructure. Experts also recommended establishing a government-run, lab-funded safety research organisation tasked with imposing safety guardrails for models exceeding a specific capability threshold.

There was also support for mandatory impact assessments and making these publicly available, as well as using tools such as a transparency index to help users understand the risks associated with a given model. Several experts advocated for introducing strict fines and penalties for non-compliance and holding vendors liable for damages to enforce accountability. Additionally, experts suggested mandatory reporting of dual-use capabilities to relevant government agencies and creating incident response plans that outline roles and responsibilities in handling potential threats.

Other suggested measures included enforcing much stricter data protection to limit available training data and reduce risks. Localised risk assessments and cross-border data-sharing agreements were recommended to address region-specific vulnerabilities and manage shared risks more effectively. Some experts also highlighted the importance of civic consultation processes to anticipate democratic risks, particularly in pre-deployment phases, ensuring that AI deployment aligns with public interests.

White-box access to government evaluators was suggested to ensure transparency, particularly for dual-use capability assessments. Additionally, experts emphasised the need for adaptive governance frameworks capable of quickly responding to emerging AI risks, especially in areas such as biosecurity and public health. They also suggested establishing AI ethics committees with diverse regional representation to address concerns from various communities, including vulnerable populations.

Several experts highlighted the need for A/B testing of models in active use, similar to social media recommender systems, to evaluate the impact of models in real-world settings. Taxing tech companies to fund external safety efforts, real-time monitoring of model performance, and creating industry incentives for independent third-party assessors were also proposed to strengthen systemic risk management. Some experts recommended restricting open access models above certain risk thresholds, and issuing post hoc penalties for systemic failures, such as financial market disruptions caused by AI.

\section{Discussion}
Our study provides a systematic evaluation of risk mitigation measures for the largest general-purpose AI models, focusing on the perceived effectiveness of measures in reducing systemic risks. By synthesising expert opinions from various domains, we have identified key measures that are considered most effective and technically feasible. In this section, we interpret these findings, discuss their implications, and acknowledge limitations.

\subsection{Interpretation of findings}
\textit{Safety incident reports and security information sharing} emerged as one of the most highly regarded measure across all expert groups. This suggests a strong consensus on the importance of transparency and collaboration in identifying and addressing AI-related incidents. \textit{Third-party pre-deployment model audits} and \textit{pre-deployment risk assessments} were also ranked highly, underscoring the value experts place on external scrutiny and proactive evaluation before AI models are deployed. The emphasis on \textit{whistleblower protections }indicates recognition of the critical role that insiders can play in identifying risks that may not be apparent through formal assessments. Interestingly, measures that rely solely on the AI providers' internal processes, such as \textit{safety vs. capabilities investments} and \textit{risk-focused governance structures}, were considered less effective unless they included independent oversight. This reflects a concern that self-regulation may be insufficient due to potential conflicts of interest.

The measures with the highest ratings are deemed very feasible, and these measures also overlap with emerging best practices at leading companies' existing voluntary commitments. Initial commitments given to the White House \citep{whitehouse2023} included external red-teaming, information sharing regarding safety risks, incentivising third party discovery of vulnerabilities, and investments in AI safety research. These commitments were re-affirmed in Seoul, where the world’s top sixteen leading AI companies further committed to set out responsible scaling type safety frameworks incorporating intolerable risk thresholds in advance of the Paris AI Summit \citep{ukgov2024}. The importance of whistleblower protections is also reflected in existing company commitments \citep{anthropic2024a,openai2024c} and other regulatory initiatives such as the SB 1047 bill that passed through the Senate in California but was eventually vetoed by the governor \citep{california2024}. Finally, some recent general-purpose AI models such as o1 by OpenAI have been submitted for external pre-deployment red teaming, indicating that some measures are at least partially being implemented and experimented with \citep{openai2024b}.

\subsection{Implications for policy}
\label{sec:4.2}
The following list presents our recommended measures. The study identifies eight priority measures for providers of general-purpose AI based on two criteria: how often experts selected them in their top-10 combinations and how strongly experts agreed with their effectiveness across all risk domains. These recommendations are grounded in expert consensus on feasibility, with all measures deemed feasible by at least 97\% of experts.

\begin{enumerate}
    \item \textbf{Third-party pre-deployment model audits:} Independent safety assessments of models before deployment, with auditors given appropriate access for testing
    \item     \textbf{Safety incident reporting and security information sharing:} Disclosure of AI incidents, near-misses, and security threats to relevant stakeholders
    \item \textbf{Whistleblower protections:} Policies ensuring safe reporting of concerns without retaliation or restrictive agreements
    \item     \textbf{Pre-deployment risk assessments:} Comprehensive assessment of potential misuse and dangerous capabilities before deployment
    \item \textbf{Risk-focused governance structures:} Implementation of board risk committees, chief risk officers, multi-party authorisation requirements, ethics boards, and internal audit teams
    \item \textbf{Intolerable risk thresholds:} Clear red lines for risk or model capabilities set by a third-party that trigger immediate development or deployment halt
    \item \textbf{Input and output filtering:} Monitoring for dangerous inputs and outputs
    \item \textbf{External assessment of testing procedure:} Third-party evaluation of how companies test for dangerous capabilities
\end{enumerate}

These measures were perceived to be some of the most effective through both selection frequency in recommended combinations and individual expert ratings. Notably, third party pre-deployment model audits, safety incident reports and security information sharing, and pre-deployment risk assessments showed particularly strong consensus across different expert groups. All these measures were selected in experts' top-10 combinations by more than 40\% of the respondents in \hyperref[fig:combinations]{Figure 5} and show over 50\% agreement combining "Strongly agree" and "Agree" in \hyperref[fig:risk1]{Figure 3} and \hyperref[fig:risk2]{Figure 4}. While all eight measures demonstrate strong overall support, their perceived effectiveness varies somewhat across different risk domains, suggesting that specific risks may benefit from different combinations of these measures.

These findings have several implications for policymakers and providers of general-purpose AI. The high rankings of measures involving external assessment and transparency suggests that regulatory frameworks should mandate independent audits and robust reporting mechanisms. Implementing safety incident reporting and security information sharing can facilitate a collective understanding of emerging risks and enable coordinated responses. The perceived effectiveness of whistleblower protections implies that legal safeguards should be established to encourage reporting of incidents of non-compliance and unethical practices without fear of retaliation. This could enhance the detection of risks that might otherwise remain hidden within organisations. 

\subsection{Strengths}
This study offers several notable strengths in its methodological approach to evaluating the perceived effectiveness of mitigation measures for systemic risks from general-purpose AI. The research design combines quantitative ratings with extensive qualitative insights from around 120 pages of expert input, allowing for richer contextual understanding. The domain-specific analysis enabled experts to evaluate measures within their areas of expertise, while allowing a structured evaluation across multiple risk categories. Future research can benefit from this qualitative data for additional insights.

The sample quality represents another key strength, featuring highly qualified participants with most having relevant publications and being in applicable senior positions. The expert panel represents diverse expertise across five critical domains to systemic risks and multiple different sectors, with good geographic distribution across the EU, US, and UK.

The survey design demonstrates comprehensiveness through its evaluation of 27 risk mitigation measures. By incorporating both individual and combinatorial effectiveness assessments, alongside feasibility ratings, the study provides insights into both theoretical effectiveness and practical implementation challenges. The qualitative rationales behind expert judgments offer valuable context for understanding the quantitative findings.

The study is highly policy relevant, contributing directly to EU AI Act implementation guidance due in May 2025, alongside other international AI governance efforts. The findings provide actionable general recommendations that align with measures already being voluntarily adopted by leading companies, bridging the gap between academic research and industry practice. The policy implications make the research valuable for policymakers developing AI governance frameworks.

Finally, we offer a transparent discussion of sample characteristics and potential biases. At the end of the paper, a large subset of experts who participated in the survey are acknowledged with permission. The clear documentation in the Appendices provides academic communities with ample opportunities for discussing our findings and interpretation as well as building upon this research further.

\subsection{Limitations}
Several limitations should be acknowledged. First, the survey relied on expert opinions, which, while valuable, may be subject to biases based on the experts' backgrounds and experiences. The sample, although diverse, may not fully represent all perspectives within the AI safety and risk domains. Notable expert groups are missing from our target list, including experts in the impact of AI on creative industries, labour markets, and education. While we strived to represent diverse risk categories and expert perspectives in the survey, the breadth of potential AI impacts necessitated selective sampling to maintain a manageable survey length and ensure quality responses.

Second, the measures were evaluated based on perceived effectiveness and feasibility, assuming they are legally required, well-executed, and overseen by competent regulators. In practice, implementation challenges may reduce their effectiveness. The assumption of optimal execution may not hold in real-world scenarios, where resource constraints, regulatory capacity and organisational cultures vary. Several participants expressed concerns that the effectiveness of the measures heavily depended on significant "ifs," such as their implementation and verification. This reliance on optimal conditions for effectiveness, which may not always be realistic, suggests that the survey may have overestimated the practical impact of some measures. The challenges of real-world implementation and the variability in difficulty across different measures were not fully captured in the survey, potentially leading to an overly optimistic assessment of their feasibility.

Third, feedback from participants suggested that providing more detailed scenarios matching risk-mitigation measures with specific threats, such as state actors using AI to develop novel pathogens or non-state actors fine-tuning models for malicious purposes, would have enhanced the relevance and precision of responses. The lack of scenario-specific contexts may have limited participants' ability to assess accurately the applicability and effectiveness of each mitigation measure across different risk categories.

Fourth, this study focuses mainly on the perceived effectiveness of various measures in mitigating general-purpose AI related systemic risks, without considering fully their associated costs or broader implications. The agreement percentages reported should not necessarily be interpreted as direct endorsements of implementation. For instance, a high percentage of experts agreeing that a particular measure could mitigate a systemic risk does not necessarily indicate their support for its adoption. This study's scope is limited to assessing potential effectiveness, and further research is required to evaluate the full spectrum of considerations, including cost-benefit analyses and feasibility. We did, however, receive a lot of qualitative input in terms of support and reservations for the measures as well as the feasibility of implementing the measures from the perspective of scientific or technical understanding of them.

Fifth, the response rate may have been affected by several factors. The survey was conducted over the holiday season in August, which likely contributed to a lower participation rate. Additionally, some individuals in the sample were very senior, which might have limited their availability to complete a lengthy survey. The survey itself was also long and demanding, and its wide scope may have made some experts feel they were not fully equipped to address all aspects comprehensively. This is evidenced by the fact that 29 respondents began the survey but did not finish it. We did not include any partial responses in the analysis. These factors likely contributed to the smaller response rate observed. Furthermore, we cannot rule out that those who chose to complete the survey versus those who did not may hold particular biases or perspectives, potentially affecting the findings.

\subsection{Future research}
Future research on expert opinion of effective mitigation measures to reduce systemic risks from general-purpose AI could aim for larger and more diverse sample sizes and analyse them from different perspectives. While expert opinion provides valuable insights, there is a critical need for empirical evidence on the effectiveness of these measures – including quantitative data on changes in incident rates, implementation costs, and concrete safety outcomes.

To address open questions related to the real-world implementation of the measures identified in this study, future research efforts should focus on the practical application and scalability of these measures. In particular, future research could explore the following key areas where further technical understanding would be beneficial: capability restrictions, risk thresholds, input/output filtering mechanisms for AI models, and information security.

Research should identify critical AI capabilities that need restriction, establish justifications linked to potential harms, and develop frameworks to assess trade-offs between economic benefits and systemic risk mitigation. Defining intolerable risk thresholds and creating adaptive models that consider evolving AI capabilities and societal contexts are also essential.

Further research is needed on adaptive input/output filtering mechanisms to prevent harmful content while balancing effectiveness with concerns about censorship. Advanced cybersecurity measures are crucial for protecting sensitive model information, focusing on insider threat safeguards and evaluating trade-offs between security and collaboration. Additionally, developing metrics to assess the effectiveness of these measures is important for ensuring responsible AI governance.

\section{Conclusion}
This study provides the first comprehensive exploration of expert perspectives on risk mitigation measures for general-purpose AI models, with a particular focus on reducing systemic risks as defined by the EU AI Act. Through a literature review and survey of 76 experts across five key areas – AI safety; critical infrastructure and democratic processes; chemical, biological, radiological, and nuclear (CBRN) risks; discrimination and bias – we identified several measures for mitigating systemic risks from general-purpose AI that are perceived to be effective and technically feasible.

Our findings reveal some expert consensus on the importance of transparency, external scrutiny, and proactive risk assessment. The most highly-rated measures include safety incident reports and security information sharing, third-party pre-deployment model audits, and pre-deployment risk assessments. Notably, experts considered all evaluated measures to be technically feasible, with even the lowest-rated measures achieving at least 91\% feasibility ratings.

Our research makes several important contributions to the field of AI governance. Firstly, we provide the first systematic assessment of risk mitigation measures that combines quantitative ratings with extensive qualitative expert insights, offering a more nuanced understanding than previous work. Secondly, our domain-specific analysis across four systemic risk areas offers insights into how different types of experts view and prioritise various mitigation measures for different risks. Thirdly, we evaluate both individual effectiveness and combinatorial impact of different measures, acknowledging the complex nature of AI risks. Finally, our findings directly inform ongoing policy discussions, particularly the implementation of the EU AI Act, by providing evidence-based recommendations for effective risk mitigation.

As the development of general-purpose AI continues to accelerate, the implementation of effective risk mitigation measures becomes increasingly critical. Our research provides a contribution for evidence-based policymaking and industry practices, while highlighting the importance of continued research and adaptation of mitigation measures as AI technology evolves.

\section{Acknowledgements}
We would like to thank Jonas Schuett, Laura Weidinger, Markus Anderljung, Mauritz Kelchtermans, and Seth Lazar for reviewing our paper and providing extensive feedback. We would also like to thank the Center for Human-Compatible AI Workshop 2024 participants who participated in our early survey and workshop.

We would like to express our sincere gratitude to all 76 respondents who participated in our survey. Among the participants, the following 57 individuals agreed to have their names and affiliations listed. Survey respondents do not represent their affiliated organisations, and the views expressed in this paper are those of the authors.

\begin{enumerate}
    \item Alice Saltini, European Leadership Network, James Martin Center for Nonproliferation Studies
    \item Alistair Knott, Victoria University of Wellington
    \item Ana Bárbara Gomes Pereira, Institute for Research on Internet and Society
    \item Angelina Wang, Stanford University
    \item Annette Zimmermann, University of Wisconsin-Madison
    \item Anthony M. Barrett, UC Berkeley
    \item Atte Ojanen, Demos Helsinki, University of Turku
    \item Ben Bucknall, Centre for the Governance of AI
    \item Benjamin Mueller, SecureBio, MIT
    \item Benjamin Plaut, University of California, Berkeley
    \item Caleb Lucas, Indiana University
    \item Caspar Oesterheld, Carnegie Mellon University
    \item Cassidy Nelson, The Centre for Long-Term Resilience
    \item Connor Dunlop, Ada Lovelace Institute
    \item Daniel S. Schiff, Purdue University, Governance and Responsible AI Lab
    \item David Manheim, The Association For Long Term Existence And Resilience (ALTER)
    \item Elizabeth Seger, Demos
    \item Elliot Jones, Ada Lovelace Institute
    \item Emily Soice, SecureBio, MIT
    \item Erdem Biyik, University of Southern California
    \item Giulio Corsi, University of Cambridge
    \item Guillem Bas, Observatorio de Riesgos Catastróficos Globales (ORCG)
    \item Hadrien Pouget, Carnegie Endowment for International Peace
    \item Hannah Rose Kirk, University of Oxford
    \item Haydn Belfield, Leverhulme Centre for the Future of Intelligence, Cambridge University
    \item Helena Puig Larrauri, Build Up
    \item Henrik Skaug Sætra, Department of Informatics, University of Oslo
    \item Henry Papadatos, SaferAI
    \item Hinda Haned, Owls \& Arrows
    \item Iason Gabriel, Google DeepMind
    \item Joe O'Brien, Institute for AI Policy \& Strategy
    \item Julia Rafal-Baer, ILO Group and Women Leading Ed
    \item Kevin Klyman, Stanford University
    \item Kevin Wei, RAND
    \item Koen Holtman, Holtman Systems Research, AI Standards Lab
    \item Lennart Heim, RAND
    \item Lorenzo Pacchiardi, University of Cambridge
    \item Madhulika Srikumar, Partnership on AI
    \item Malcolm Murray, SaferAI
    \item Marta Ziosi, University of Oxford
    \item Micah Musser, New York University
    \item Michael Frank, Seldon Strategies, 2430 Group
    \item Michelle Loi, Algorithmwatch.org
    \item Peter Babigumira Ahabwe, Nuo Bioscience
    \item Richard Moulange, Centre for Long-Term Resilience
    \item Roland Pihlakas, independent AI safety researcher
    \item Salvatore Romano, AI Forensics
    \item Sarah Carter, Science Policy Consulting
    \item Sarah Kreps, Cornell University
    \item Sella Nevo, RAND Corporation
    \item Shelby Grossman, Stanford Internet Observatory
    \item Stephen Casper, MIT
    \item Tan Zhi-Xuan, MIT
    \item Tatiana Caldas-Löttiger, International WomenX in Business for AI Ethics
    \item Theodore Lechterman, IE University
    \item Thomas H. Costello, MIT
    \item Veronica Stefan, Digital Citizens Romania
\end{enumerate} 

\newpage
\bibliography{references}

%%%%%%%%%%%%%%%%%%%%%%%%%%%%%%%%%%%%%%%%%%%%%%%%%%%%%%%%%%%%
\newpage

\appendix

\section{Appendices}
\subsection{CHAI Conference survey and workshop}
\label{appendix:A}
At an early stage of this research, we hosted a workshop at the Center for Human-Compatible AI Conference in California on the 14th of June 2024. Prior to the workshop, participants completed a survey in which they rated 37 risk mitigation measures based on their perceived effectiveness in reducing system risks from general-purpose AI. Participants were instructed to imagine a comprehensive and well-executed version of each mitigation measure when providing their evaluations. The survey included the following prompt to guide participants' assessments: "Rate the following measures in terms of how effective they are likely to be at reducing systemic risks from general-purpose AI in the next five years." Participants were instructed to consider how each measure could reduce total risk, conceptualised as Risk = Likelihood x Severity of impact. It was clarified that for a mitigation measure to be considered 'extremely effective,' it does not need to eliminate all risk on its own.

Participants represented a range of affiliations, including MIT, UC Berkeley, Carnegie Mellon University, and Google DeepMind. Data collected from the surveys were analysed to identify trends and consensus on the most effective measures for reducing AI-related systemic risks.

The list of risk mitigation measures presented to participants was the following:

\textbf{Intolerable risk thresholds}: Establish red-line risk or capability thresholds, continuously assess and monitor for breaches, and prepare technical, legal, and organisational measures to immediately halt development and deployment if a breach occurs (\href{https://cltc.berkeley.edu/publication/ai-risk-management-standards-profile}{Barrett et al., 2023}, p. 32, 54; \href{https://assets.publishing.service.gov.uk/media/653aabbd80884d000df71bdc/emerging-processes-frontier-ai-safety.pdf}{UK Government, 2023}, p. 11-12; \href{https://doi.org/10.48550/arXiv.2310.00328}{O'Brien et al., 2023}, p. 20, 24).

\textbf{Safety vs. capabilities}: A significant fraction of employees working on enhancing model safety and alignment rather than capabilities (\href{https://cdn.governance.ai/AGI_Safety_Governance_Practices_GovAIReport.pdf}{Schuett et al., 2023}, p. 11, 18).

\textbf{Pre-deployment risk assessments}: Comprehensive risk assessments before deployment that would assess foreseeable misuse and include dangerous capability evaluations, incorporating post-training enhancements and collaborations with domain experts (\href{https://cdn.governance.ai/AGI_Safety_Governance_Practices_GovAIReport.pdf}{Schuett et al., 2023}, p. 12, 18; \href{https://partnershiponai.org/wp-content/uploads/1923/10/PAI-Model-Deployment-Guidance.pdf}{Partnership on AI, 2023}, p. 3; \href{https://doi.org/10.48550/arXiv.2404.02675}{Kolt et al., 2024}, p. 4; \href{https://doi.org/10.48550/arXiv.2305.15324}{Shevlane et al., 2023}, p. 9; \href{https://cltc.berkeley.edu/publication/ai-risk-management-standards-profile}{Barrett et al., 2023}, p. 21; \href{https://assets.publishing.service.gov.uk/media/653aabbd80884d000df71bdc/emerging-processes-frontier-ai-safety.pdf}{UK Government, 2023}, p. 17).

\textbf{Third-party model audits}: External assessment to provide a judgement — or input to a judgement — on the safety of a model. Audits might be conducted by governments or other third-parties (\href{https://www.whitehouse.gov/wp-content/uploads/2023/07/Ensuring-Safe-Secure-and-Trustworthy-AI.pdf}{The White House, 2023}, p. 3; \href{https://partnershiponai.org/wp-content/uploads/1923/10/PAI-Model-Deployment-Guidance.pdf}{Partnership on AI, 2023}, p. 3; \href{https://cdn.governance.ai/AGI_Safety_Governance_Practices_GovAIReport.pdf}{Schuett et al., 2023}, p. 18; \href{https://assets.publishing.service.gov.uk/media/653aabbd80884d000df71bdc/emerging-processes-frontier-ai-safety.pdf}{UK Government, 2023}, p. 17).

\textbf{Whistleblower protections}: Comprehensive whistleblower protection policies that clearly outline relevant reporting processes, protection mechanisms, and non-retaliation assurances while refraining from restrictive non-disparagement agreements (\href{https://cdn.governance.ai/AGI_Safety_Governance_Practices_GovAIReport.pdf}{Schuett et al., 2023}, p. 21; \href{https://cltc.berkeley.edu/publication/ai-risk-management-standards-profile}{Barrett et al., 2023}, p. 24).

\textbf{Deploying powerful models in stages}: Starting with a small number of applications and fewer users, and gradually scaling up as confidence in the model’s safety increases (\href{https://cdn.governance.ai/AGI_Safety_Governance_Practices_GovAIReport.pdf}{Schuett et al., 2023}, p. 19; \href{https://cltc.berkeley.edu/publication/ai-risk-management-standards-profile}{Barrett et al., 2023}, p. 59, 76).

\textbf{Advanced information security}: Implementing advanced cybersecurity measures and insider threat safeguards to protect proprietary and unreleased model weights (\href{https://assets.publishing.service.gov.uk/media/653aabbd80884d000df71bdc/emerging-processes-frontier-ai-safety.pdf}{UK Government, 2023}, p. 24; \href{https://cltc.berkeley.edu/publication/ai-risk-management-standards-profile}{Barrett et al., 2023}, p. 47; \href{https://www.whitehouse.gov/wp-content/uploads/2023/07/Ensuring-Safe-Secure-and-Trustworthy-AI.pdf}{The White House, 2023}, p. 3; \href{https://cdn.governance.ai/AGI_Safety_Governance_Practices_GovAIReport.pdf}{Schuett et al., 2023}, p. 12, 18; \href{https://www-cdn.anthropic.com/de8ba9b01c9ab7cbabf5c33b80b7bbc618857627/Model_Card_Claude_3.pdf}{Anthropic, 2024a}, p. 26; \href{https://doi.org/10.48550/arXiv.2305.15324}{Shevlane et al., 2023}, p. 9).

\begin{enumerate}
    \item \textbf{Prohibiting high-stakes applications}: Enforcing use policies that prohibit high-stakes applications. Requires Know-Your-Customer procedures  (\href{https://doi.org/10.48550/arXiv.2310.00328}{O'Brien et al., 2023}, p. 10, 12; \href{https://cdn.governance.ai/AGI_Safety_Governance_Practices_GovAIReport.pdf}{Schuett et al., 2023}, p. 19).

    \item \textbf{Safety incident reports}: Disclosing concrete harms, safety incidents and near misses to appropriate state actors and other providers (e.g. via an AI incident database) (\href{https://assets.publishing.service.gov.uk/media/653aabbd80884d000df71bdc/emerging-processes-frontier-ai-safety.pdf}{UK Government, 2023}, p. 20; \href{https://partnershiponai.org/wp-content/uploads/1923/10/PAI-Model-Deployment-Guidance.pdf}{Partnership on AI, 2023}, p. 4; \href{https://doi.org/10.48550/arXiv.2404.02675}{Kolt et al., 2024}, p. 7; \href{https://doi.org/10.48550/arXiv.2310.00328}{O'Brien et al., 2023}, p. 24; \href{https://cdn.governance.ai/AGI_Safety_Governance_Practices_GovAIReport.pdf}{Schuett et al., 2023}, p. 19; \href{https://doi.org/10.48550/arXiv.2305.15324}{Shevlane et al., 2023}, p. 9).
\end{enumerate}

\textbf{Advanced model access for vetted external researchers}: Examples of advanced access rights could include any of the following: increased control over sampling, access to fine-tuning functionality, the ability to inspect and modify model internals, access to additional information like used training data, or additional features like stable model versions with back-compatibility (\href{https://assets.publishing.service.gov.uk/media/653aabbd80884d000df71bdc/emerging-processes-frontier-ai-safety.pdf}{UK Government, 2023}, p. 18, \href{https://cdn.governance.ai/Structured_Access_for_Third-Party_Research.pdf}{Bucknall \& Trager, 2024}, p. 12).

\textbf{Vetted researcher access}: Giving good faith, public interest evaluation researchers access to research APIs that provide technical and legal safe harbours to limit barriers imposed by usage policy enforcement and potential legal repercussions associated with stringent terms of service (\href{https://doi.org/10.48550/arXiv.2305.15324}{Shevlane et al., 2023}, p. 6, \href{https://assets.publishing.service.gov.uk/media/653aabbd80884d000df71bdc/emerging-processes-frontier-ai-safety.pdf}{UK Government, 2023}, p. 36; \href{https://cdn.governance.ai/AGI_Safety_Governance_Practices_GovAIReport.pdf}{Schuett et al., 2023}, p. 19; \href{https://arxiv.org/pdf/2403.04893}{Longpre et al., 2024, p. 7}).

\textbf{Preventing capability jumps}: Not deploying models that are far more capable than any existing models. This implies increasing the amount of compute spent training frontier models only incrementally (e.g., not by more than three times between each increment) (\href{https://cdn.governance.ai/AGI_Safety_Governance_Practices_GovAIReport.pdf}{Schuett et al., 2023}, p. 20; \href{https://www-cdn.anthropic.com/1adf000c8f675958c2ee23805d91aaade1cd4613/responsible-scaling-policy.pdf}{Anthropic, 2023}, p. 12).

\textbf{Security information sharing}: Sharing threat intelligence and information related to security incidents with appropriate government(s) and industry stakeholders (\href{https://assets.publishing.service.gov.uk/media/653aabbd80884d000df71bdc/emerging-processes-frontier-ai-safety.pdf}{UK Government, 2023}, p. 20; \href{https://www.whitehouse.gov/wp-content/uploads/2023/07/Ensuring-Safe-Secure-and-Trustworthy-AI.pdf}{The White House, 2023}, p. 2; \href{https://cdn.governance.ai/AGI_Safety_Governance_Practices_GovAIReport.pdf}{Schuett et al., 2023}, p. 18; \href{https://doi.org/10.48550/arXiv.2404.02675}{Kolt et al., 2024}, p. 4).

\textbf{Bug bounty programs}: Clear and user-friendly bug bounty programs that acknowledge and reward individuals for reporting unknown vulnerabilities and dangerous capabilities (\href{https://cltc.berkeley.edu/publication/ai-risk-management-standards-profile}{Barrett et al., 2023}, p. 27, 48; \href{https://cdn.governance.ai/AGI_Safety_Governance_Practices_GovAIReport.pdf}{Schuett et al., 2023}, p. 18; \href{https://assets.publishing.service.gov.uk/media/653aabbd80884d000df71bdc/emerging-processes-frontier-ai-safety.pdf}{UK Government, 2023}, p. 4, 29; \href{https://www.whitehouse.gov/wp-content/uploads/2023/07/Ensuring-Safe-Secure-and-Trustworthy-AI.pdf}{The White House, 2023}, p. 3).

\textbf{Publishing risk assessments}: Publishing results of internal and external pre-deployment risk assessments, except when this would unduly reveal proprietary information or itself produce significant risk. More detailed technical information about safety evaluations can be shared with the appropriate government(s) (\href{https://cdn.governance.ai/AGI_Safety_Governance_Practices_GovAIReport.pdf}{Schuett et al., 2023}, p. 19; \href{https://cltc.berkeley.edu/publication/ai-risk-management-standards-profile}{Barrett et al., 2023}, p. 19; \href{https://assets.publishing.service.gov.uk/media/653aabbd80884d000df71bdc/emerging-processes-frontier-ai-safety.pdf}{UK Government, 2023}, p. 13; \href{https://partnershiponai.org/wp-content/uploads/1923/10/PAI-Model-Deployment-Guidance.pdf}{Partnership on AI, 2023}, p. 3; \href{https://doi.org/10.48550/arXiv.2305.15324}{Shevlane et al., 2023}, p. 9).

\textbf{Capability restrictions}: Restricting risky capabilities of deployed models, such as advanced autonomy (e.g., self-assigning new sub-goals, executing long-horizon tasks) or tool use functionalities (e.g., function calls, web browsing) (\href{https://doi.org/10.48550/arXiv.2310.00328}{O'Brien et al., 2023}, p. 12, 13).

\textbf{Input and output filtering}: Monitoring for dangerous outputs (e.g., code that appears to be malware, viral genome sequences) and inputs that violate acceptable use policies to ensure models do not engage in harmful behaviour (\href{https://doi.org/10.48550/arXiv.2310.00328}{O'Brien et al., 2023}, p. 12, 24; \href{https://assets.publishing.service.gov.uk/media/653aabbd80884d000df71bdc/emerging-processes-frontier-ai-safety.pdf}{UK Government, 2023}, p. 38).

\textbf{KYC screenings}: Know-your-customer (KYC) screenings before granting access to models with very high misuse potential or to users producing large amounts of output (\href{https://cdn.governance.ai/AGI_Safety_Governance_Practices_GovAIReport.pdf}{Schuett et al., 2023}, p. 19; \href{https://assets.publishing.service.gov.uk/media/653aabbd80884d000df71bdc/emerging-processes-frontier-ai-safety.pdf}{UK Government, 2023}, p. 39).

\textbf{Staged release prior to open sourcing}: A staged-release approach before publicly releasing the weights of powerful models. Only releasing model weights if no substantial risks or harms have emerged over a sufficient time period after a release via API (with red teaming and other evaluations as appropriate) (\href{https://cltc.berkeley.edu/publication/ai-risk-management-standards-profile}{Barrett et al., 2023}, p. 59).

\textbf{Safety drills}: Regularly practising the implementation of an emergency response plan to stress test the organisation’s ability to respond to foreseeable, fast-moving emergency scenarios (\href{https://cdn.governance.ai/AGI_Safety_Governance_Practices_GovAIReport.pdf}{Schuett et al., 2023}, p. 19; \href{https://cltc.berkeley.edu/publication/ai-risk-management-standards-profile}{Barrett et al., 2023}, p. 60).

\textbf{Pre-registration of large training runs}: Registering upcoming training runs above a certain size with an appropriate state actor. Such reports could include descriptions of architecture, training compute, data collection and curation, training objectives and techniques, and planned risk management procedures (\href{https://cdn.governance.ai/AGI_Safety_Governance_Practices_GovAIReport.pdf}{Schuett et al., 2023}, p. 20).

\textbf{Fine-tuning restrictions}: Restricting or closely monitoring fine-tuning of models to ensure safeguards remain intact (\href{https://doi.org/10.48550/arXiv.2310.00328}{O'Brien et al., 2023}, p. 12).

\textbf{Adversarial robustness}: State-of-the-art methods such as adversarial training to make models robust to adversarial attacks (e.g., jailbreaking) (\href{https://cltc.berkeley.edu/publication/ai-risk-management-standards-profile}{Barrett et al., 2023}, p. 18; \href{https://arxiv.org/pdf/2312.11805}{Gemini Team et al., 2024}, p. 36).

\textbf{Transparent governance structures}: Publicly disclosing how high-stakes decisions regarding model development and deployment are made (\href{https://cdn.governance.ai/AGI_Safety_Governance_Practices_GovAIReport.pdf}{Schuett et al., 2023}, p. 19).

\textbf{Data curation}: Careful data curation prior to all development stages (including fine-tuning) to filter out high-risk content and ensure the training data is sufficiently high-quality (\href{https://arxiv.org/pdf/2312.11805}{Gemini Team et al., 2024}, p. 28; \href{https://cltc.berkeley.edu/publication/ai-risk-management-standards-profile}{Barrett et al., 2023}, p. 29, 56; \href{https://assets.publishing.service.gov.uk/media/653aabbd80884d000df71bdc/emerging-processes-frontier-ai-safety.pdf}{UK Government, 2023}, p. 43).

\textbf{Pre-development risk assessments}: Conducting risk assessments based on forecasted capabilities before training new models (\href{https://www-cdn.anthropic.com/1adf000c8f675958c2ee23805d91aaade1cd4613/responsible-scaling-policy.pdf}{Anthropic, 2023}, p. 13; \href{https://assets.publishing.service.gov.uk/media/653aabbd80884d000df71bdc/emerging-processes-frontier-ai-safety.pdf}{UK Government, 2023}, p. 10; \href{https://doi.org/10.48550/arXiv.2305.15324}{Shevlane et al., 2023}, p. 9; \href{https://arxiv.org/pdf/2303.08774}{OpenAI, 2023b}, p. 59).

\textbf{Harmlessness training}: State-of-the-art reinforcement learning and fine-tuning techniques to ensure models do not engage in unsafe behaviour (\href{https://assets.publishing.service.gov.uk/media/653aabbd80884d000df71bdc/emerging-processes-frontier-ai-safety.pdf}{UK Government, 2023}, p. 39; \href{https://www-cdn.anthropic.com/de8ba9b01c9ab7cbabf5c33b80b7bbc618857627/Model_Card_Claude_3.pdf}{Anthropic, 2024a}, p. 10; \href{https://arxiv.org/pdf/2312.11805}{Gemini Team et al., 2024}, p. 29; \href{https://arxiv.org/pdf/2303.08774}{OpenAI, 2023b}, p. 61).

\textbf{Internal audits}: An internal audit team that reports directly to the board of directors, operates independently of development teams, and is tasked with assessing the adequacy of the company's risk management practices (\href{https://partnershiponai.org/wp-content/uploads/1923/10/PAI-Model-Deployment-Guidance.pdf}{Partnership on AI, 2023}, p. 3; \href{https://cdn.governance.ai/AGI_Safety_Governance_Practices_GovAIReport.pdf}{Schuett et al., 2023}, p. 19; \href{https://cltc.berkeley.edu/publication/ai-risk-management-standards-profile}{Barrett et al., 2023}, p. 22, 40; \href{https://assets.publishing.service.gov.uk/media/653aabbd80884d000df71bdc/emerging-processes-frontier-ai-safety.pdf}{UK Government, 2023}, p. 14; \href{https://www.whitehouse.gov/wp-content/uploads/2023/07/Ensuring-Safe-Secure-and-Trustworthy-AI.pdf}{The White House, 2023}, p. 2; \href{https://doi.org/10.48550/arXiv.2404.02675}{Kolt et al., 2024}, p. 4; \href{https://doi.org/10.48550/arXiv.2305.15324}{Shevlane et al., 2023}, p. 6).

\textbf{Unlearning techniques}: Removing specific harmful capabilities (e.g., pathogen design) from models using unlearning techniques \href{https://arxiv.org/pdf/2403.03218}{(Li et al., 2024)}.

\textbf{Input data audits}: Third-party assessments that ensure all training data is free of high-risk content (\href{https://partnershiponai.org/wp-content/uploads/1923/10/PAI-Model-Deployment-Guidance.pdf}{Partnership on AI, 2023}, p. 4; \href{https://assets.publishing.service.gov.uk/media/653aabbd80884d000df71bdc/emerging-processes-frontier-ai-safety.pdf}{UK Government, 2023}, p. 42-43).

\textbf{Chief risk officers}: Senior executives who are responsible for risk management (\href{https://cdn.governance.ai/AGI_Safety_Governance_Practices_GovAIReport.pdf}{Schuett et al., 2023}, p. 19; \href{https://cltc.berkeley.edu/publication/ai-risk-management-standards-profile}{Barrett et al., 2023}, p. 22).

\textbf{Board risk committee}: A permanent committee within the board of directors that oversees the company’s risk management practices (\href{https://cdn.governance.ai/AGI_Safety_Governance_Practices_GovAIReport.pdf}{Schuett et al., 2023}, p. 19; \href{https://cltc.berkeley.edu/publication/ai-risk-management-standards-profile}{Barrett et al., 2023}, p. 22).

\textbf{Ethics reviews}: A dedicated internal body or process (e.g., an ethics board) tasked with reviewing development and deployment decisions (\href{https://cltc.berkeley.edu/publication/ai-risk-management-standards-profile}{Barrett et al., 2023}, p. 22).

\textbf{User access restrictions}: Models that automatically impose access restrictions on users based on historical misuse. Possible restrictions include banning users or throttling their model access (e.g., reduced context windows, access frequency limits, or restrictions on fine-tuning access) (\href{https://doi.org/10.48550/arXiv.2310.00328}{O'Brien et al., 2023}, p. 11).

\textbf{Dual control}: Critical decisions in model development and deployment (e.g. promotion to production, changes to training datasets, or modifications to production) made by a minimum of two individuals (\href{https://cdn.governance.ai/AGI_Safety_Governance_Practices_GovAIReport.pdf}{Schuett et al., 2023}, p. 19).

\textbf{Deepfake restrictions}: Not enabling users to create deepfakes of specific individuals (\href{https://www.schumer.senate.gov/imo/media/doc/Roadmap_Electronic1.32pm.pdf}{Schumer et al., 2024}).

\textbf{Identifiers of AI-generated content}: State-of-the-art methods for watermarking AI outputs and identifying AI-generated content (\href{https://partnershiponai.org/wp-content/uploads/1923/10/PAI-Model-Deployment-Guidance.pdf}{Partnership on AI, 2023}, p. 4; \href{https://cltc.berkeley.edu/publication/ai-risk-management-standards-profile}{Barrett et al., 2023}, p. 18; \href{https://assets.publishing.service.gov.uk/media/653aabbd80884d000df71bdc/emerging-processes-frontier-ai-safety.pdf}{UK Government, 2023}, p. 32; \href{https://www.whitehouse.gov/wp-content/uploads/2023/07/Ensuring-Safe-Secure-and-Trustworthy-AI.pdf}{The White House, 2023}, p. 3).

\subsubsection{Survey and workshop results}

The following section first presents the qualitative and then the quantitative results of our survey and workshop. The qualitative results consist of some of the rationales experts gave to justify their ratings. The quantitative section summarises the effectiveness-ratings of different mitigation measures.

\paragraph{Quantitative results.}

\begin{figure}
    \centering
    \includegraphics[width=\linewidth]{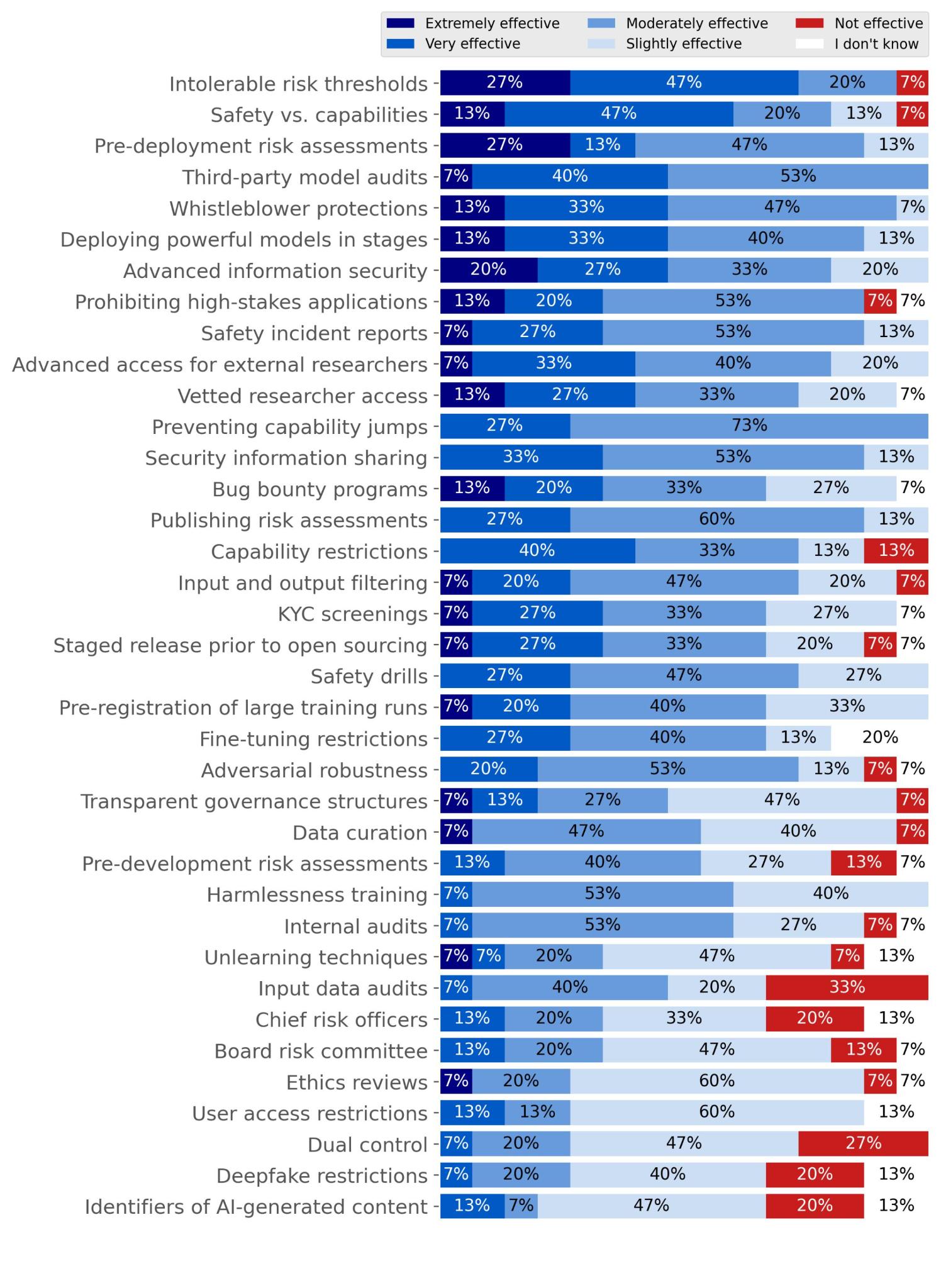}
    \caption{Risk mitigation measures, ordered from most effective to least effective, as rated by experts. n=15.}
    \label{fig:chai}
\end{figure}

\hyperref[fig:chai]{Figure 7} summarises the results of the survey on risk mitigation measures. The bars represent the percentage of participants who rated each measure at specific levels of effectiveness, with different colours indicating varying levels of agreement. These quantitative insights provide an overview of expert consensus regarding the perceived effectiveness of each proposed mitigation measure.

\paragraph{Qualitative results.} 
For the five highest and the five lowest rated risk mitigation measures, we highlight some of the written comments experts contributed. Furthermore, we highlight some general remarks made in the in-depth workshop discussions.

Highest-rated measures:

1. \textbf{Intolerable risk thresholds}: “Effective if identified by legitimate 3rd parties”

2. \textbf{Safety vs capabilities}: “A significant increase of effort on safety vs capabilities would greatly improve the current situation where capabilities are far better resourced”

3. \textbf{Pre-deployment risk assessments}: “Of course this should happen and many should start thinking about the risks of next models already after the last generation was deployed”

4. \textbf{Third-party model audits}: “Only powerful if acceptable risk is defined and models that fail will be prevented from deployment”

5. \textbf{Whistleblower protections}: “Great for cutting across many interventions, and helpful for especially those leaving companies, or even protection for external investigators”

Lowest-rated measures:

1.\textbf{ Identifiers of AI-generated content}: “Narrowly useful but bad actors have open source alternatives”

2. \textbf{Deepfake restrictions}: “The serious threats come from actors who can evade these and these may create a false sense of security”

3. \textbf{Dual control}: “Can be circumvented if two individuals not properly vetted”

4. \textbf{User access restrictions}: “Too easy to open new accounts”

5. \textbf{Ethics reviews}: “Recent events at OpenAI indicate how ineffective these can be”

In the workshop discussions attendees highlighted that no single risk mitigation measure is sufficient to address systemic risks; instead, a portfolio of complementary measures is necessary. They emphasised that effective implementation depends on specificity and empirical evidence, stressing that well-designed measures must be diligently executed to succeed. Concerns were also raised about relying solely on providers of general-purpose AI for risk management, as self-regulation could compromise effectiveness due to conflicts of interest. Independent oversight and transparency were identified as essential components for overcoming these challenges and ensuring effective risk mitigation.

\subsubsection{Key takeaways from the CHAI survey and workshop}
The following conclusions have been drawn from this preliminary analysis of risk mitigation measures:

\begin{enumerate}
    \item No single risk mitigation measure is effective on its own, a portfolio of approaches is needed. None of the mitigation measures are rated as \textit{extremely effective} by a majority of the experts. To sufficiently reduce systemic risk, a selection of risk mitigation measures is required.
\item More specifics and empirical information are needed in order to make risk mitigation effective. Detailed, context-specific data and empirical studies are necessary to understand the nuances of implementing effective risk mitigation strategies.
\item Much of the effectiveness of risk mitigations depends on how well they are executed.
\item Proper implementation and diligent execution of risk mitigation measures are essential for their success. Even well-designed measures can fail if not properly executed.
\item Many interventions become ineffective if controlled solely by the providers with self-interests.Mitigation measures should not be exclusively managed by general-purpose AI providers, as their self-interests may compromise the effectiveness. Independent oversight and involvement of third parties are essential.
\item Some measures are likely ineffective due to information asymmetries. Discrepancies in information and knowledge among stakeholders can lead to ineffective implementation of risk mitigation measures. Ensuring transparency and information sharing is important to overcome these asymmetries.
\end{enumerate}

\subsection{Survey questions}
\label{appendix:B}
This appendix presents the primary contents of our survey, including the introduction presented to participants and the key questions they were asked to respond to.

\subsubsection{Introduction}
The following introduction was provided to survey participants:

“In the following questions, you will be asked whether the \textbf{implementation of different risk mitigation measures by providers of large general-purpose AI models would effectively reduce systemic risks}. When considering a specific mitigation measure, you should assume that the measure would be \textbf{legally required, well-executed, and overseen by a competent regulator.}

This survey focuses on the \textbf{large general-purpose AI models}, specifically the ones which the EU AI Act classifies as having systemic risks. Under the EU AI Act, general-purpose AI models are considered to have systemic risks when the cumulative amount of computation used for its training measured in floating point operations exceeds 10\^25 FLOP.

Currently, the general-purpose AI models in scope likely include \textbf{GPT-4o, Gemini Ultra 1.5, Claude 3.5, and Llama 3.1 405B}, and possibly others.”

\subsubsection{Systemic risks}
The following definition and categorisation of systemic risks were provided to participants:

“The EU AI Act defines systemic risk as "a risk that is specific to the high-impact capabilities of general-purpose AI models, having a significant impact on the Union market due to their reach, or due to actual or reasonably foreseeable negative effects on public health, safety, public security, fundamental rights, or the society as a whole, that can be propagated at scale across the value chain."

We divide the general-purpose AI-related systemic risks mentioned in the AI Act into the following four categories:

\begin{enumerate}
    \item \textbf{Disruptions of critical sectors}, such as digital infrastructure or other critical infrastructure, including via cyberattacks;
    \item \textbf{Negative effects on democratic processes}, such as civil discourse and electoral procedures;
    \item \textbf{Chemical, biological, radiological, and nuclear risks}, and other serious risks to public health and safety;
    \item \textbf{Harmful bias and discrimination} with risks to communities or societies.”
\end{enumerate}

\subsubsection{Risk reduction}
The following instructions on risk reduction were provided:

“We are interested in identifying the risk mitigation measures that most effectively reduce systemic risks.

Reducing the risk means reducing the total systemic risk, where ‘risk’ is defined as the combination of the probability of an occurrence of harm and the severity of that harm.

Additional instructions

\begin{itemize}
    \item If you are unfamiliar with a particular measure or otherwise feel like you cannot make an informed judgement, please select "\textbf{I don't know}".
\end{itemize}

If you believe it is not feasible to implement a certain measure robustly because the scientific or technical understanding of the intervention is not mature enough, please choose “\textbf{not feasible}” and expand on your rationale in a comment.”

\subsubsection{Main questions}
The following key questions were presented to participants:

\textbf{a) “To what extent do you agree or disagree that the implementation of this risk mitigation measure by providers of large general-purpose AI models would effectively reduce the following systemic risks?} Assume the measure is legally required, well-executed, and overseen by a competent regulator.

Provide your assessment for each systemic risk separately. You are invited to elaborate on your reasoning for each of the measures in the box below.

\begin{itemize}
    \item Additional Information
    \begin{itemize}
        \item Reducing the risk means reducing the total systemic risk where ‘risk’ is defined as the combination of the probability of an occurrence of harm and the severity of that harm.
        \item If you are unfamiliar with a particular measure or otherwise feel like you cannot make an informed judgement, please select “\textbf{I don’t know}”.
        \item If you believe it is not feasible to robustly implement a certain measure because the scientific or technical understanding of the intervention is not mature enough, please select "\textbf{not feasible}" and mention your rationale as a comment.”
    \end{itemize}
\end{itemize}
Note: To answer this question, participants were presented with a table containing the four systemic risk categories: (1) Disruptions of critical sectors; (2) Negative effects on democratic processes; (3) Chemical, biological, radiological, and nuclear (CBRN) risks; and (4) Harmful bias and discrimination. Participants could select from the following responses: Strongly disagree; Somewhat disagree; Neutral; Somewhat agree; Strongly agree; I don’t know; Not feasible.

\textbf{b) “What are the reasons for your previous response? }Feel free to use bullet points for brevity.” 

\textbf{c) “Please select the combination of 10 measures which you estimate would be most effective at reducing the systemic risks from general-purpose AI. }Assume each measure is legally required, well-executed, and overseen by a competent regulator.”

\textbf{d) “What are the reasons for your previous response? }Feel free to use bullet points for brevity.”

\textbf{e) “Are there other mitigation measures that a regulator could require providers of large general-purpose AI models to implement which you think would be effective at reducing the systemic risks from general-purpose AI? }Assume the measure is legally required, well-executed, and overseen by a competent regulator. You can write your response briefly and in bullet points to save time if you wish to do so.”

\subsection{Qualitative expert insights per measure}
\label{appendix:C}
For each measure, we append summaries of the qualitative input provided by the experts in the survey.

\textbf{Pre-development risk assessments}

Expert opinions on pre-development risk assessments for AI models varied, but many agreed that this measure could be helpful as a best practice, particularly for detecting potential emergent capabilities and assessing risks early in the design phase. Several experts emphasised its importance in guiding responsible development decisions and mitigating risks before they materialise. However, multiple experts expressed scepticism about the accuracy of capability forecasting before development, noting the difficulty in predicting emergent capabilities and the limitations of current forecasting methods.

The effectiveness of these assessments, according to many experts, would heavily depend on who conducts them and their access to data. Several stressed the need for independence in conducting these assessments, suggesting involvement from government and independent experts rather than relying solely on tech industry actors. Some experts highlighted the need for expertise beyond AI, including social sciences and economics, to evaluate potential risks comprehensively.

Regarding specific risk areas, several experts noted that this measure could be particularly beneficial for chemical, biological, radiological, nuclear (CBRN) risks; and critical infrastructure disruptions. Its potential usefulness in addressing bias, discrimination, and democratic process risks was also mentioned by multiple experts, though some noted the challenges in predicting these impacts theoretically. While some experts considered pre-development assessments more effective than post-deployment ones, others emphasised the need for ongoing risk assessment throughout the AI lifecycle. Several experts suggested this measure might be more crucial for frontier models or systems with autonomous capabilities. Some proposed improvements, such as using small models to inform forecasts (scaling-informed forecasts) and establishing clear processes, frameworks, and standards for conducting these assessments.

Concerns were raised by some experts about the potential for this measure to become a form of "security theatre" or "assessment-washing" without proper implementation. Several worried it might slow down innovation or be used as justification to train potentially dangerous models. Some experts also pointed out that this approach might advantage large firms with substantial compliance budgets.

\textbf{Pre-deployment risk assessments}

Multiple experts agreed that pre-deployment risk assessments could be helpful across all risk areas as a general best practice. They noted that such assessments allow for proactive identification and mitigation of potential risks before AI models are deployed. Several experts emphasised the importance of involving domain experts in the assessment process to ensure thorough and informed evaluations.

Some experts cautioned that the effectiveness of these assessments would depend heavily on who performs them and their level of independence. Several mentioned that internal assessments might be biased towards deployment due to conflicts of interest. Several experts pointed out that pre-deployment assessments may be more effective for certain risks (like chemical, biological, radiological, and nuclear (CBRN) risks; and critical infrastructure disruption) that are more concrete and easier to test for, compared to more diffuse risks like impacts on democratic processes.

Several experts noted that while helpful, these assessments have limitations. They mentioned the difficulty in predicting all potential risks, especially for unforeseen or evolving threats. Some suggested that ongoing monitoring and iterative assessments would be necessary to complement pre-deployment evaluations. Individual experts raised additional points, such as the need for standardisation, the potential for false negatives, and the importance of these assessments being substantive rather than mere box-ticking exercises.

\textbf{Third party pre-deployment model audits}

Multiple experts agreed that third-party audits would be helpful across all risk areas as a general best practice. They noted that external evaluations could catch issues before deployment and provide valuable independent perspectives. However, several experts also highlighted challenges, including the current lack of standardised evaluation methods for many risks, potential for flawed or biased audits, and limitations of pre-deployment testing to predict real-world impacts.

Some experts emphasised the importance of auditor independence and expertise. Several suggested that governments may not be the ideal auditors due to potential lack of AI expertise, with independent third-party experts seen as potentially more effective.

Individual experts made various additional points:

\begin{itemize}
    \item Audits may be most useful for specific, well-defined risks like bias or chemical, biological, radiological, and nuclear (CBRN) risks.
    \item The process could be burdensome and slow innovation.
    \item Combining audits with ongoing monitoring and adaptive measures is important.
    \item There's a need for clear audit criteria and standardised methodologies.
    \item Audits alone may not be sufficient for complex, long-term, or systemic risks.

\end{itemize}

\textbf{External assessment of testing procedure}

Most experts viewed this measure positively overall, with many highlighting its potential to provide unbiased perspectives, improve risk identification, and enhance safety across various risk categories. Several experts mentioned that external assessments could catch issues internal teams might miss. Multiple experts emphasised the importance of red-teaming exercises as part of these assessments.

However, numerous caveats and concerns were raised. Multiple experts questioned the current capabilities and expertise of external evaluation firms, particularly for specialised domains like chemical, biological, radiological, and nuclear (CBRN) risks. Several noted that the effectiveness would depend heavily on the quality and independence of the evaluators. Several experts expressed concern about potential conflicts of interest or the risk of evaluations becoming mere "rubber stamps."

Some experts highlighted that this approach might be more effective for certain types of risks (e.g., cybersecurity) than others (e.g., broader societal impacts). Several mentioned that while helpful, external evaluations might not capture all deployment risks. Several experts suggested that this measure should complement, not replace, internal evaluation capabilities. Several recommended government or multi-stakeholder oversight of the evaluation process. Single experts raised various other points, including the need for clear evaluation criteria, the potential for this to create a new commercial sector similar to financial auditing, and the importance of ensuring that evaluation results are acted upon.

\textbf{Vetted researcher access}

Many experts viewed this measure as generally beneficial for risk reduction across multiple areas. Several noted its potential to identify new risks and threat vectors, particularly for bias and discrimination, and chemical, biological, radiological, and nuclear (CBRN) risks. Multiple experts highlighted the value of independent scrutiny and collective intelligence in uncovering potential issues.

However, concerns were raised by multiple experts about implementation challenges. These included determining who qualifies as a "good faith" researcher, managing potential conflicts of interest, and balancing transparency with information security risks. Several experts pointed out that API access alone may be limiting, suggesting full model access could be more effective.

Some experts emphasised the measure's particular usefulness for bias and discrimination research, as it could allow minority groups to test models without significant financial barriers. Others noted it may be less effective for more diffuse risks like impacts on democratic processes. A minority of experts expressed scepticism about the measure's impact, citing limitations of black-box access or questioning whether it would significantly help with certain risk types like biosecurity.

\textbf{Advanced model access for vetted external researchers}

Multiple experts viewed this measure positively, seeing it as beneficial for identifying and mitigating risks across various domains. They highlighted its potential for enhancing oversight, improving safety evaluations, and fostering innovation. Several experts emphasised the importance of having diverse perspectives and independent auditing to counterbalance corporate interests.

Some experts cautioned about implementation challenges, including proper vetting processes, managing potential conflicts of interest, and ensuring information security. Several noted that while helpful, this measure alone may not be sufficient to guarantee safety or fully mitigate all risks.

Individual experts raised specific points:

\begin{itemize}
    \item One suggested it could be particularly useful for addressing bias issues.
    \item Another emphasised its value for "red-teaming" exercises.
    \item One expert noted its potential to reduce information hazards associated with certain AI capabilities.
    \item A single expert expressed concern about companies potentially using this for "safety-washing."

\end{itemize}

\textbf{Data curation}

Multiple experts agreed that this measure could be effective for reducing bias and discrimination risks, with some noting it as a fundamental or critical step. Several experts also saw potential benefits for mitigating chemical, biological, radiological and nuclear (CBRN) risks, particularly by filtering out specialised data. However, opinions were mixed on its effectiveness for other risk areas. Some experts expressed scepticism about its impact on democratic process risks and critical infrastructure disruption. Several noted the difficulty in defining and identifying "high-risk" content in these domains.

Several experts highlighted implementation challenges, including the vast amount of data involved, potential reduction in model capabilities, and the difficulty of thorough curation without significantly limiting the training set. Several mentioned that companies likely already attempt some level of data curation for performance reasons. Some experts cautioned that this approach alone is insufficient, as models may still be able to recreate dangerous knowledge from "safe" data. One expert noted that careful data curation might be more effective for near-term models (in the next two years) than for more advanced future systems.

Several experts emphasised the importance of balancing risk mitigation with maintaining useful capabilities, noting the dual-use nature of much scientific and technical information. One expert argued against the concept of "high-quality" data, suggesting all data could be useful depending on the application.

\textbf{Harmlessness training}

Multiple experts noted that harmlessness training techniques like RLHF and DPO can be helpful in reducing certain risks, particularly those related to bias, discrimination, and harmful language. However, many also emphasised that these methods are not sufficient on their own and have limitations. Several experts pointed out that harmlessness training can be bypassed or "jailbroken" by determined actors, making it less effective against high-impact risks or threats from malicious users. Several mentioned that it may be more effective against novice threats than sophisticated ones.

Some experts raised concerns about defining "unsafe behaviour" or "harm," noting the lack of consensus and potential cultural differences. A couple suggested that oversight of implementation would be necessary. Multiple experts commented that harmlessness training might be more effective for everyday interactions but less so for extreme cases or specific incidents related to critical infrastructure or chemical, biological, radiological, and nuclear (CBRN) risks.

Several experts mentioned that these techniques might create a superficial "mask" over the model's behaviour without fundamentally changing its capabilities or knowledge. Some individual points raised included: the need to consider the wellbeing of workers involved in the training process, the potential for these techniques to become outdated, and the importance of balancing multiple objectives rather than focusing on single values.

\textbf{Adversarial robustness}

Multiple experts agreed that adversarial robustness training is generally helpful across various risk areas and should be considered a best practice. However, opinions on its effectiveness varied. Several experts noted that while it is a useful tool, it may not be sufficient on its own to address all risks comprehensively. Some experts highlighted that current methods might not be robust enough against determined adversaries or sophisticated jailbreak attempts. Several mentioned that the effectiveness of these techniques can vary over time and may require continuous updates.

There was a consensus among several experts that this approach could be more effective for certain types of risks (e.g., hate speech, bias) than others (e.g., impacts on democratic processes). Some pointed out that it might be less effective for addressing systemic or subtle effects arising from AI use in specific contexts. Several experts emphasised the importance of integrating adversarial robustness with other safety measures and ongoing monitoring. One expert suggested that governmental disclosure and reporting obligations should accompany such measures.

Some individual experts raised concerns about potential limitations, such as the risk of overfitting to specific attack types, the challenge of achieving true adversarial robustness in foundation models, and the possibility that robust safeguards might inadvertently limit the model's capabilities for benign tasks.

\textbf{Unlearning techniques}

Multiple experts expressed scepticism about the effectiveness and feasibility of unlearning techniques. Several noted that these techniques are nascent, not well understood, and may not be robust in the long term. Several experts mentioned that unlearning might be circumvented, especially with open-source models or through fine-tuning.

Several experts suggested that unlearning could be more effective for specific, concrete risks like chemical, biological, radiological and nuclear (CBRN) risks, or cybersecurity threats, where particular knowledge can be targeted. However, multiple experts doubted its efficacy for broader issues like bias, discrimination, or impacts on democratic processes, which are more complex and less clearly defined. Several experts raised concerns about potential unintended consequences, such as creating knowledge gaps or affecting the model's overall performance. One expert noted that capabilities are not cleanly separable, making targeted unlearning challenging.

Some experts saw potential benefits, with one mentioning compliance with laws and regulations, and another suggesting it could enhance model security. However, these positive views were in the minority. Multiple experts emphasised the need for more research and a focus on evaluating outcomes rather than specific techniques. Several suggested combining unlearning with other mitigation strategies and oversight measures.

\textbf{Deploying powerful models in stages}

Multiple experts agreed that staged deployment could be beneficial for early risk detection and management before full release. They noted it allows for monitoring real-world effects on a smaller scale, helps identify unforeseen risks, and provides time to adapt mitigations. Several experts mentioned this approach is common in other industries and software development.

However, multiple experts also pointed out limitations. Some argued it may not effectively capture risks that only emerge at larger scales or with more diverse users. Others noted that bad actors could simply wait for full deployment before attempting serious misuse. Several experts highlighted that this approach would not apply to open-source models and might be less useful if the model will eventually be open-sourced anyway. One expert mentioned it could create inequality by restricting initial access.

Individual experts raised various other points, including: the need to couple this with other measures, the importance of proper information gathering during stages, potential challenges in defining stages, and the risk of creating false assurance if major risks are missed.

\textbf{Fine-tuning restrictions}

Multiple experts agreed that fine-tuning restrictions could be helpful in reducing risks, particularly for bias mitigation and as a general model security practice. Several noted it could significantly reduce the threat space and preserve safeguards. However, there were mixed opinions on its effectiveness across different risk areas.

Some experts pointed out limitations, such as:

\begin{itemize}
    \item The measure may not be feasible for open-source models
    \item Existing methods to bypass restrictions may make fine-tuning unnecessary
    \item Uncertainty about how well companies can detect dangerous fine-tuning data
    \item Potential to hinder legitimate improvements and beneficial uses
\end{itemize}
Several experts emphasised the importance of monitoring and accountability in the fine-tuning process. One suggested implementing ethical reviews as part of fine-tuning. Several experts expressed uncertainty about the measure's effectiveness or feasibility, citing a lack of research or understanding of fine-tuning's impacts on removing safeguards.

Individual experts raised various points, including:

\begin{itemize}
    \item The measure might be more effective for specific risks (e.g., biodesign tools) rather than general-purpose AI models
    \item It could be less useful for addressing bias and discrimination risks arising from benign use cases
    \item The importance of balancing restrictions with the potential benefits of fine-tuning

\end{itemize}

\textbf{Capability restrictions}

Multiple experts agreed that capability restrictions could be effective, particularly for chemical, biological, radiological, and nuclear (CBRN) risks; and critical infrastructure disruptions. They noted that limiting advanced autonomy and tool use functionalities could significantly reduce these risks. However, several experts also expressed concerns about the feasibility of implementing such restrictions, citing economic incentives and the difficulty of isolating specific capabilities.

Some experts mentioned that capability restrictions might be less effective for addressing bias and discrimination risks or negative effects on democratic processes, as these issues are not necessarily tied to advanced capabilities. Several experts pointed out that restricting capabilities could potentially drive development underground or be easily circumvented by skilled programmers. One expert emphasised the importance of maintaining human oversight and control, while another suggested that restrictions should be risk-scenario based rather than function-based. A single expert raised concerns about potential censorship and over-politicisation of AI if restrictions are too broad.

Several experts noted the challenge of balancing risk mitigation with preserving beneficial uses of AI, particularly in research and economic applications. The need for international agreements and enforcement was mentioned by one expert.

\textbf{KYC screenings}

Multiple experts suggested that KYC screenings could be more effective for specific, tangible risks like chemical, biological, radiological, and nuclear (CBRN) risks; and critical infrastructure disruptions; rather than for addressing bias or impacts on democratic processes. Several noted that KYC might help reduce deliberate misuse and narrow the field of potential adversaries, particularly for less sophisticated threat actors.

However, many experts expressed scepticism about KYC's overall effectiveness. Common concerns included:

\begin{itemize}
    \item Difficulty in implementation, especially for open-source or locally-run models.
    \item Limited utility against state actors or advanced threats who can circumvent screenings.
    \item Potential privacy violations and exclusion of certain user communities.
    \item Ineffectiveness against unintentional or distributed harms.
\end{itemize}
Some experts highlighted KYC's use in other industries (like finance) as a positive precedent, while others pointed out its limitations even in those contexts. Several mentioned that KYC could be part of a broader risk management strategy but should not be relied upon solely. Individual experts raised additional points, such as the need for tiered access systems, concerns about hampering legitimate research, and suggestions for combining KYC with other measures like real-time use monitoring.

\textbf{Prohibiting high-stakes applications}

Multiple experts agreed that prohibiting high-stakes applications could be effective for reducing risks in critical infrastructure, chemical, biological, radiological, and nuclear (CBRN) risk scenarios; and some specific use cases. However, several noted that this approach may be less effective for addressing bias and democratic process concerns, as these risks are more diffuse and harder to define.

A common concern raised by multiple experts was the difficulty of enforcement, particularly for open-weight models or against determined adversaries. Several mentioned that malicious actors might easily circumvent such prohibitions.

Some experts highlighted the need for clearer definitions of what constitutes "high-stakes" applications, as this term can be interpreted differently across various contexts. Several suggested that prohibitions should be balanced with the potential benefits of AI in certain high-stakes areas, such as healthcare.

Individual experts raised additional points, including:

\begin{itemize}
    \item The potential for prohibitions to reduce scrutiny and safety standards in non-prohibited areas
    \item The importance of government support in enforcing such policies
    \item Concerns about privacy implications of Know-Your-Customer (KYC) procedures
    \item The need for flexible legislation that allows for safe AI deployment if rigorous standards are met
    \item The potential for prohibitions to drive development into more obscure or less regulated areas

\end{itemize}

\textbf{Input and output filtering}

Multiple experts noted that input/output filtering could be effective for specific risks like chemical, biological, radiological, and nuclear (CBRN) risks; and cybersecurity issues, where harmful content is more clearly definable. However, several also pointed out that sophisticated actors might still find ways to circumvent these filters.

Many experts expressed scepticism about the effectiveness of filtering for more complex or nuanced risks like bias, discrimination, and threats to democratic processes. They argued these issues are harder to define and detect through simple filtering mechanisms.

Several experts mentioned that while filtering can be useful, it has limitations. Some noted it may be more effective for average users but less so for determined malicious actors. Others pointed out that overzealous filtering could potentially infringe on free speech or hinder legitimate uses.

Several experts highlighted the importance of monitoring inputs and outputs for identifying misuse patterns and improving risk mitigation strategies over time. However, concerns about privacy and data protection were also raised. Some experts emphasised the need for clear definitions of harmful content and transparent policies governing the filtering process. A couple mentioned the challenge of keeping filters up-to-date with evolving threats.

\textbf{Bug bounty programs}

Multiple experts noted that bug bounty programs can be effective and have proven useful in other areas like cybersecurity. They are seen as a cost-effective way to leverage crowdsourced knowledge and promote accountability. Several experts mentioned that these programs could be helpful across various risk areas, particularly for identifying biases, discrimination, and vulnerabilities that could negatively affect democratic processes.

However, there were also several concerns raised by multiple experts:

\begin{itemize}
    \item The effectiveness may be limited for certain types of risks, especially those related to critical infrastructure or chemical, biological, radiological, and nuclear (CBRN) risks.
    \item There are potential information hazards, as the process might inadvertently teach people how to circumvent AI safeguards.
    \item The programs might not be as effective for subjective or value-laden issues.
    \item They occur post-deployment, which may be too late for some critical risks.
\end{itemize}
Some individual experts raised specific points:

\begin{itemize}
    \item One suggested that bug bounties could create a false sense of security if not properly implemented.
    \item Another mentioned that in certain contexts (like Brazil), such programs could be vulnerable to scams.
    \item One expert pointed out that these programs should not substitute robust, mandated risk assessments and expert auditing.

\end{itemize}

\textbf{Safety drills}

Multiple experts noted that safety drills would be most effective for acute, fast-moving scenarios, particularly those involving critical infrastructure, cybersecurity or chemical, biological, radiological, and nuclear (CBRN) risks. Several mentioned that this approach would be less useful for slower-developing issues like bias and impacts on democratic processes.

Several experts highlighted the value of such drills in defining roles, responsibilities, and improving organisational preparedness. Some drew parallels to similar practices in other high-risk industries. Several experts expressed uncertainty about how to define or effectively implement these drills for AI-specific scenarios. A couple mentioned that regulators might lack the necessary expertise to oversee such plans effectively.

Some individual experts raised points about the measure being potentially "too little, too late" for certain risks, the importance of including independent oversight, and the need to integrate these drills with broader risk management practices. Several experts suggested that while helpful, this measure alone would not be sufficient to address all types of AI risks and should be part of a more comprehensive approach to AI safety and governance.

\textbf{Intolerable risk thresholds}

Multiple experts agreed that risk thresholds could be helpful, especially for chemical, biological, radiological, and nuclear (CBRN) risks; and infrastructure risks. However, many also noted challenges in implementation. Several experts mentioned difficulties in defining and operationalising appropriate thresholds, particularly for more abstract risks like bias or effects on democratic processes.

Several experts emphasised the importance of third-party evaluation and legally defined red lines. Some suggested this approach could build useful norms and mechanisms within AI companies. Multiple experts expressed scepticism about companies actually halting development if thresholds were breached, citing competitive pressures. Several noted that risk thresholds might be more effective for certain types of risks (e.g. chemical, biological, radiological, and nuclear (CBRN) risks) than others (e.g. bias).

Several experts highlighted the need for ongoing research and flexibility in setting thresholds, as our understanding of AI risks evolves. Several mentioned that thresholds alone are insufficient and should be part of a broader regulatory approach. Some individual experts raised specific points, such as the potential for thresholds to become bureaucratic box-ticking exercises, the challenge of applying thresholds to open-source models, and the possibility that the most effective offensive systems might also be the best defensive ones.

\textbf{Risk-focused governance structures}

Many experts viewed this measure positively, seeing it as helpful across various risk areas. Multiple experts emphasised that while important, these structures need to be implemented effectively to have real impact. Several noted that the quality and effectiveness could vary significantly between companies.

Several experts highlighted that such governance is a necessary foundation but may be insufficient on its own. Some suggested it should be combined with other measures like strong safety culture or external oversight. Multiple experts pointed out that these structures may be more effective for concrete risks like cybersecurity or chemical, biological, radiological, and nuclear (CBRN) risks; and less so for complex, systemic risks like impacts on democratic processes.

Several experts raised concerns about potential downsides, including creating bureaucracy, slowing innovation, or providing a false sense of security. Several were sceptical about real-world effectiveness, with one citing experiences in finance where such structures were often gamed or ineffective. Some individual points raised included: the need for these structures to "have teeth" with robust enforcement, the importance of independence and external reporting capabilities for oversight boards, and the suggestion to adopt a "three lines of defence" model for risk management.

\textbf{Whistleblower protections}

Multiple experts agreed that whistleblower protections are generally beneficial and important across all risk areas. They noted that such protections can enhance transparency, allow for quicker identification of issues, and serve as an early warning system for potential risks. Several experts mentioned that whistleblowers have historically been crucial in exposing harmful activities in various industries.

Some experts highlighted that whistleblower protections might be particularly effective for risks related to bias, discrimination and misinformation, Several noted that these protections could be less effective for risks to democratic processes or critical sectors, as these may be harder to identify or more diffuse. Multiple experts emphasised that whistleblower protections should be mandated by regulations rather than relying on voluntary corporate commitments. Several also pointed out the need to balance these protections with safeguards for trade secrets and to prevent potential abuse.

Several experts expressed scepticism about the practical effectiveness of such protections, citing past instances of retaliation against whistleblowers in non-AI contexts. One expert noted potential conflicts with intellectual property rights. Some individual points raised included: the need to extend protections to third-party contractors and evaluators, the importance of creating safe reporting channels to minimise information hazards, and the suggestion to include whistleblower protections as part of a broader package including safe harbours and bug bounty programs.

\textbf{Safety vs. capabilities investments}

Multiple experts agreed this measure could be broadly helpful in reducing risks across various areas. However, several noted challenges in implementation and effectiveness. Common concerns included difficulty defining "safety" vs "capabilities" work, potential for companies to game the system, and questions about feasibility.

Some experts highlighted the importance of not just quantity but quality of safety investments. Several mentioned this could be particularly impactful for issues like discrimination and bias. Multiple experts stressed the need for external oversight or incentives to ensure effectiveness.

Individual experts raised various points: one suggested embedding safety throughout the organisation rather than in separate teams; another noted potential unintended consequences favouring large tech companies; and one highlighted the tension between safety investments and market competitiveness.

\textbf{Safety incident reports and security information sharing}

Many experts viewed this measure positively overall, with several noting its importance across different risk areas. Multiple experts highlighted its value for addressing security-related risks like chemical, biological, radiological, and nuclear (CBRN) risks; and critical infrastructure disruptions. Some saw it as less effective for bias and democratic process risks. Several experts drew parallels to successful reporting systems in other industries like aviation. The importance of timely reporting and including "near-misses" was mentioned by multiple experts. Several experts emphasised the need for clear processes and incentives around reporting. The potential for information overload was noted by one expert. 

Some individual points raised include:

\begin{itemize}
    \item The need to balance transparency with national security concerns for WMD-related incidents
    \item The importance of sharing information with both governments and other AI companies
    \item Potential challenges in defining reportable "incidents" for some risk types
    \item The value of using this data to inform regulation and future risk mitigation
\end{itemize}
A couple of experts cautioned that while useful, this measure alone is not sufficient and should be combined with other preventative approaches. One expert noted it may be more effective for discrete harms rather than cumulative effects.

\textbf{Sharing safety cases}

Multiple experts noted that sharing safety cases could enhance transparency, accountability and regulatory oversight. However, several also expressed concerns about the potential for companies to abuse the "proprietary information" exception to avoid full disclosure. Several experts mentioned that this measure could be particularly effective for more concrete risks like chemical, biological, radiological, and nuclear (CBRN) risks; and critical infrastructure disruptions, but less so for complex societal issues like impacts on democratic processes.

Some experts worried that safety cases might become a "box-checking exercise" or ineffective bureaucratic requirement without proper enforcement or incentives for thorough evaluation. A couple of experts suggested that safety cases could be valuable if combined with other measures like third-party audits or red-teaming exercises. Individual experts raised various other points, including: the need for a robust safety culture within organisations, the potential for safety cases to incentivise more thorough internal evaluations, and the importance of balancing transparency with protection of proprietary information.

\textbf{Transparent governance structures}

Multiple experts agreed that transparency in governance structures is generally beneficial and important for accountability, but opinions varied on its effectiveness in directly mitigating specific risks. Several experts noted that while helpful, transparency alone is not sufficient and must be coupled with other measures. A common view was that transparency could be particularly useful for understanding and mitigating infrastructure risks and risks related to democratic processes. However, multiple experts expressed uncertainty about how regulators would effectively use or act on the disclosed information.

Some experts highlighted potential challenges, including the difficulty of ensuring full and meaningful disclosures, the potential burden on smaller companies, and the need to balance transparency with confidentiality and trade secrets. Several experts mentioned that transparency could help standardise best practices across the industry and prepare governments for potential risks. However, one expert cautioned that relying solely on company disclosures might create a false sense of security.

\textbf{Pre-registration of large training runs}

Multiple experts viewed pre-registration as potentially helpful for improving transparency, allowing government oversight, and enabling proactive risk assessment. However, several also expressed concerns about its limitations. Some noted that pre-registration alone may not directly reduce risks without accompanying enforcement powers or interventions. Several experts mentioned that it could help prepare mitigations and prevent strategic surprises.

Several experts questioned the measure's effectiveness for specific systemic risks like bias or democratic process disruption, suggesting these issues are not necessarily tied to model size or training run scale. A couple pointed out that pre-registration might be more relevant for existential risks or "lab accidents" than the listed systemic risks. Concerns raised by individual experts included: potential to discourage experimentation, difficulty in predicting outcomes from training descriptions alone, risk of triggering "race dynamics" if information is made public, and significant compliance costs for companies. One expert worried about the potential for abuse in a "licence to compute" system.

Some experts suggested the indirect effects could be valuable, such as improving regulators' knowledge and setting groundwork for further regulation. Several noted the importance of international enforcement for effectiveness.

\textbf{Advanced information security}

Multiple experts agreed that this measure is generally helpful across all risk areas as a security best practice. Several noted its particular importance for protecting against misuse by malicious actors, including state-level threats. Several experts highlighted its relevance for safeguarding critical infrastructure and chemical, biological, radiological, and nuclear (CBRN) risks.

Some experts expressed uncertainty about the measure's overall impact, with one noting the lack of precedent for stealing foundation models. Several mentioned that the effectiveness depends on how much more dangerous private models are compared to publicly available ones. One expert argued it is less relevant for bias and discrimination risks, as these issues are more likely to arise from improper use of legitimate models. Another suggested it may not significantly affect misinformation risks, as weaker public models could suffice for such purposes.

Several experts viewed this as a necessary but not sufficient measure. One described it as a "bare minimum" that is unobjectionable but not particularly powerful for advancing AI safety. Another called it a "no-brainer." Several experts raised concerns about the measure's applicability in a world with open-source models, with one strongly opposing forcing open-source models to close.

\subsection{Additional risk mitigation measures suggested by experts}
\label{appendix:D}
In addition to evaluating the predefined list of risk mitigation measures, we invited participants to suggest other potential measures that regulators could require providers of large general-purpose AI models to implement. Experts were asked to consider measures that could be effective at reducing systemic risks from general-purpose AI, assuming these measures would be legally required, well-executed, and overseen by a competent regulator. This open-ended question allowed us to capture innovative ideas and perspectives that may not have been included in our initial set of measures. The following subsections present these expert-proposed measures, grouped by thematic areas.

\begin{enumerate}
    \item \textbf{\textbf{Governance and Regulatory Frameworks}}
    \begin{enumerate}
        \item Government-run, lab-funded organisation for safety research and imposing guardrails
        \item Strict fines and penalties for non-compliance
        \item Holding vendors liable for damages
        \item Adaptive governance frameworks
        \item Taxing tech companies
        \item Post hoc penalties for systemic breaches
        \item Establishing a government-funded safety institute
        \item International cooperation to close loopholes and dark sites
        \item Licensing of models and monitoring of model use
        \item Criminal liability for harms caused by non-audited AI applications
        \item Export controls
        \item IP controls to deal with labour disruption risks
        \item On-site inspections by regulators
        \item Consortium model for AGI development with multi-government representation
        \item Embedded monitors/regulators within industry actors
        \item Overarching risk management frameworks with specific action thresholds
        \item Measures ensuring digital sovereignty
        \item Fair reparations for victims of AI-caused damage
    \end{enumerate}
\end{enumerate}

\begin{enumerate}
    \item \textbf{\textbf{Transparency and Documentation}}
    \begin{enumerate}
        \item Publishing impact assessments and mitigation plans in a public repository
        \item Usage of the transparency index (MIT index) and explanation of resulting score
        \item Mandatory model attribution
        \item Publicly documenting all training data
        \item Model documentation and transparency requirements
        \item Model cards
        \item Careful documentation for downstream users and developers
        \item Providing technical data and safety reasoning to downstream deployers/users
        \item Traceability and cryptographic attribution requirements for outputs of general-purpose AI models
        \item Implementing watermarking methods for generated content
        \item Substantial transparency (e.g., foundation model transparency index)
        \item Accountability and documentation on data sources and acquisition methods
    \end{enumerate}
\end{enumerate}

\begin{enumerate}
    \item \textbf{\textbf{Safety Evaluation and Testing}}
    \begin{enumerate}
        \item Coordinated red-teaming
        \item Layered deployment with required internal/external evaluations at pre-training and pre-release stages
        \item Enforcing evaluation of possible impact on public use cases (high-stake applications)
        \item A/B tests of models in active use
        \item White-box access to government evaluators for dual-use capability assessments
        \item Red-teaming of other companies' safety systems
        \item Mandatory evaluation and public reporting of model abilities
        \item Bias/discrimination testing with pre-defined metrics compliant with non-discrimination laws
        \item Performing evaluations at several checkpoints throughout the model lifecycle
    \end{enumerate}
\end{enumerate}

\begin{enumerate}
    \item \textbf{\textbf{Ethical and Societal Considerations}}
    \begin{enumerate}
        \item Democratically legitimate civic consultation processes, especially pre-deployment
        \item Localised risk assessments for region-specific vulnerabilities
        \item Policies for ethical AI use in healthcare
        \item AI ethics committees with regional representation
        \item Consultation with external experts, researchers, communities, and vulnerable groups
        \item Interdisciplinary collaboration on ethics and democracy
        \item Justice-led frameworks considering systemic impacts of AI technologies
        \item Incorporating culture risk in ethical impact assessments
        \item Equal resources spent on each language the model supports
        \item Citizen panels or mini-publics for consulting on acceptable uses
        \item Getting informed consent from communities before deployment
        \item Analysis of specific impact risks for vulnerable groups
        \item Inclusion of diverse groups in all processes
        \item Multidisciplinary teams for evaluating AI models and their societal impacts
        \item Demonstrating AI prioritisation of universal human ethical values over business interests
    \end{enumerate}
\end{enumerate}

\begin{enumerate}
    \item \textbf{\textbf{Safety Research and Development}}
    \begin{enumerate}
        \item Percentage of revenue to fund external safety efforts
        \item Research on AI models' contribution to life sciences progress
        \item Interpretability requirements
        \item More diverse research teams
        \item Measures to subsidise independent third-party assessors and evaluators
        \item Measures to subsidise independent input/output filter modules and safety components
        \item Measures to subsidise independent authors of reward and loss functions
        \item Public infrastructure, funding, and development of AI models and safety techniques
    \end{enumerate}
\end{enumerate}

\begin{enumerate}
    \item \textbf{\textbf{Deployment and Operational Safety}}
    \begin{enumerate}
        \item Required lag between model training and model release
        \item Maintaining and providing incident response plans to regulators
        \item Post-deployment monitoring
        \item Real-time monitoring of model performance and drift
        \item Recovery and rollback plans
        \item Human-in-the-loop requirements for agents interacting with outside systems
        \item Regular updates to models based on newly discovered risks
        \item Keeping rate limits low to reduce capacity for large-scale attacks
        \item Logging user queries for identifying bad actors
        \item Restrictions or nudges on user prompts/use cases
        \item Active monitoring and termination of AI-enabled agents/worms/viruses/botnets
        \item Automatic rollbacks of model deployment if serious incidents occur
        \item Data centre monitoring/audits to prevent unpermitted frontier training runs
    \end{enumerate}
\end{enumerate}

\begin{enumerate}
    \item \textbf{\textbf{Capability Control and Restrictions}}
    \begin{enumerate}
        \item Ban on autonomous agentic AI systems using safety-tuned base models
        \item Restriction of particular new architectures designed to increase capabilities
        \item Coordinated blackout periods across organisations
        \item No open access models above a certain risk or capability threshold
        \item Authorisation requirements for accessing potentially dangerous system capabilities
        \item Banning specific domains of use (e.g., no military use)
        \item Outlawing military and police use of general-purpose AI models
        \item Disallowing irreversible release of open model weights prior to evaluation
    \end{enumerate}
\end{enumerate}

\begin{enumerate}
    \item \textbf{\textbf{Other measures}}
    \begin{enumerate}
        \item Stricter data protection and intellectual property protection
        \item Cross-border data sharing agreements
        \item AI literacy and capacity building programs
        \item Training developers in ethics and/or making them personally liable for harms
        \item Economic incentives to preserve human skills in key security areas
        \item International coordination, standardisation, and information sharing
        \item Mandated reporting of dual-use capabilities to relevant government agencies
        \item Responsible development guidelines and scaling policies
        \item Demonstrating AI observation beyond expected consequences of its actions
    \end{enumerate}
\end{enumerate}

\subsection{Initial risk mitigation measures identified through literature review}
\label{appendix:E}

\begin{enumerate}
    \item Pre-deployment risk assessment. AGI labs should take extensive measures to identify, analyse, and evaluate risks from powerful models before deploying them.
    \item Dangerous capability evaluations. AGI labs should run evaluations to assess their models’ dangerous capabilities (e.g. misuse potential, ability to manipulate, and power-seeking behaviour).
    \item Third-party model audits. AGI labs should commission third-party model audits before deploying powerful models.
    \item Safety restrictions. AGI labs should establish appropriate safety restrictions for powerful models after deployment (e.g. restrictions on who can use the model, how they can use the model, and whether the model can access the internet)
    \item Red teaming. AGI labs should commission external red teams before deploying powerful models.
    \item Monitor systems and their uses. AGI labs should closely monitor deployed systems, including how they are used and what impact they have on society.
    \item Alignment techniques. AGI labs should implement state-of-the-art safety and alignment techniques.
    \item Security incident response plan. AGI labs should have a plan for how they respond to security incidents (e.g. cyberattacks).
    \item Post-deployment evaluations. AGI labs should continually evaluate models for dangerous capabilities after deployment, taking into account new information about the model’s capabilities and how it is being used.
    \item Report safety incidents. AGI labs should report accidents and near misses to appropriate state actors and other AGI labs (e.g. via an AI incident database).
    \item Safety vs capabilities. A significant fraction of employees of AGI labs should work on enhancing model safety and alignment rather than capabilities.  
    \item Internal review before publication. Before publishing research, AGI labs should conduct an internal review to assess potential harms.
    \item Pre-training risk assessment. AGI labs should conduct a risk assessment before training powerful models.
    \item Emergency response plan. AGI labs should have and practice implementing an emergency response plan. This might include switching off systems, overriding their outputs, or restricting access.
    \item Protection against espionage. AGI labs should take adequate measures to tackle the risk of state-sponsored or industrial espionage.
    \item Pausing training of dangerous models. AGI labs should pause the development process if sufficiently dangerous capabilities are detected.
    \item Increasing level of external scrutiny. AGI labs should increase the level of external scrutiny in proportion to the capabilities of their models.
    \item Publish alignment strategy. AGI labs should publish their strategies for ensuring that their systems are safe and aligned.
    \item Bug bounty programs. AGI labs should have bug bounty programs, i.e. recognise and compensate people for reporting unknown vulnerabilities and dangerous capabilities.
    \item Industry sharing of security information. AGI labs should share threat intelligence and information about security incidents with each other.
    \item Security standards. AGI labs should comply with information security standards (e.g. ISO/IEC 27001 or NIST Cybersecurity Framework). These standards need to be tailored to an AGI context.
    \item Publish results of internal risk assessments. AGI labs should publish the results or summaries of internal risk assessments, unless this would unduly reveal proprietary information or itself produce significant risk. This should include a justification of why the lab is willing to accept remaining risks.
    \item Dual control. Critical decisions in model development and deployment should be made by at least two people (e.g. promotion to production, changes to training datasets, or modifications to production).
    \item Publish results of external scrutiny. AGI labs should publish the results or summaries of external scrutiny efforts, unless this would unduly reveal proprietary information or itself produce significant risk.
    \item Military-grade information security. The information security of AGI labs should be proportional to the capabilities of their models, eventually matching or exceeding that of intelligence agencies (e.g. sufficient to defend against nation states).
    \item Board risk committee. AGI labs should have a board risk committee, i.e. a permanent committee within the board of directors which oversees the lab’s risk management practices.
    \item Chief risk officer. AGI labs should have a chief risk officer (CRO), i.e. a senior executive who is responsible for risk management.  
    \item Statement about governance structure. AGI labs should make public statements about how they make high-stakes decisions regarding model development and deployment.
    \item Publish views about AGI risk. AGI labs should make public statements about their views on the risks and benefits from AGI, including the level of risk they are willing to take in its development.
    \item KYC screening. AGI labs should conduct know-your-customer (KYC) screenings before giving people the ability to use powerful models.
    \item Third-party governance audits. AGI labs should commission third-party audits of their governance structures.
    \item Background checks. AGI labs should perform rigorous background checks before hiring/appointing members of the board of directors, senior executives, and key employees.
    \item Model containment. AGI labs should contain models with sufficiently dangerous capabilities (e.g. via boxing or air-gapping).
    \item Staged deployment. AGI labs should deploy powerful models in stages. They should start with a small number of applications and fewer users, gradually scaling up as confidence in the model’s safety increases.
    \item Tracking model weights. AGI labs should have a system that is intended to track all copies of the weights of powerful models.
    \item Internal audit. AGI labs should have an internal audit team, i.e. a team which assesses the effectiveness of the lab’s risk management practices. This team must be organisationally independent from senior management and report directly to the board of directors.
    \item No open-sourcing. AGI labs should not open-source powerful models, unless they can demonstrate that it is sufficiently safe to do so.
    \item Researcher model access. AGI labs should give independent researchers API access to deployed models.
    \item API access to powerful models. AGI labs should strongly consider only deploying powerful models via an application programming interface (API).
    \item Avoiding hype. AGI labs should avoid releasing powerful models in a way that is likely to create hype around AGI (e.g. by overstating results or announcing them in attention-grabbing ways).  
    \item Gradual scaling. AGI labs should only gradually increase the amount of compute used for their largest training runs.
    \item Treat updates similarly to new models. AGI labs should treat significant updates to a deployed model (e.g. additional fine-tuning) similarly to its initial development and deployment. In particular, they should repeat the pre-deployment risk assessment.
    \item Pre-registration of large training runs. AGI labs should register upcoming training runs above a certain size with an appropriate state actor.  
    \item Enterprise risk management. AGI labs should implement an enterprise risk management (ERM) framework (e.g. the NIST AI Risk Management Framework or ISO 31000). This framework should be tailored to an AGI context and primarily focus on the lab’s impact on society.
    \item Treat internal deployments similarly to external deployments. AGI labs should treat internal deployments (e.g. using models for writing code) similarly to external deployments. In particular, they should perform a pre-deployment risk assessment.
    \item Notify a state actor before deployment. AGI labs should notify appropriate state actors before deploying powerful models.
    \item Notify affected parties. AGI labs should notify parties who will be negatively affected by a powerful model before deploying it.
    \item Inter-lab scrutiny. AGI labs should allow researchers from other labs to scrutinise powerful models before deployment.
    \item Avoid capabilities jumps. AGI labs should not deploy models that are much more capable than any existing models.
    \item Notify other labs. AGI labs should notify other labs before deploying powerful models.
    \item Blocklisting individuals or groups. Imposing IP or other verification-based restrictions on users based on anticipated or historical misuse.
    \item Throttle number of prompts. Place a hard cap on the number of prompts that can be submitted to a model in a given amount of time.
    \item Throttle number of calls. Place a hard cap on the number of function calls (e.g., JSON documents sent to an external API) that a single model can output in a given amount of time.
    \item Reduce context windows. Reduce the number of tokens a model is capable of processing in relation to one another. This curbs a model’s capabilities by reducing its ability to “remember” earlier information.
    \item Limit user ability to fine-tune. Frontier AI developers could remove this functionality for users, or retract specific fine-tuned instances.
    \item Output filtering. Monitor and automatically filter out dangerous outputs, such as code that appears to be malware, or viral genome sequences. 
    \item Removal of dangerous capabilities. Attempt to remove specific capabilities (e.g., pathogen design) via fine-tuning, reinforcement learning from human feedback, concept erasure, or other methods.
    \item Global planning limits. Adjust whether the same model instance has access to a large number of users, or is limited to more narrow sets of interactions.
    \item Autonomy limits. For example, restricting the ability for a model to define new actions (e.g., via assigning itself new sub-goals in an iterative loop), or to execute tasks (versus solely responding to queries).
    \item Prohibiting high-stakes applications. Setting a use policy that restricts the model from being used in high-stakes applications, and allows banning or otherwise penalising users that breach this policy. Requires Know-Your-Customer procedures.
    \item Tool use limits. Limit the ability of a model to interact with downstream tools (e.g., to use other APIs), make function calls, browse the web, etc.
    \item Make necessary preparations for rapid model rollback in emergencies.
    \item Develop an Incident response plan.
    \begin{enumerate}
        \item Instituting controls to prevent or mitigate the severity of incidents, including defining fall-backs for downstream users, especially in safety-critical domains;
        \item Establishing triggers for deployment corrections based on thresholds set and maintained as part of the threat modelling process;
        \item Developing a documented response plan for executing deployment corrections which clearly delineates roles and responsibilities;
        \item Ensuring that industry partners (such as partnering tech companies, and compute providers) adopt the above tools and protocols.
    \end{enumerate}
    \item Secure input-output monitoring
    \item Responsible Capability Scaling provides a framework for managing risk as organisations scale capability of frontier AI systems. It enables companies to prepare for potential future, more dangerous AI risks before they occur, as well as manage the risks associated with current systems. Practices involve conducting thorough risk assessments, pre-specifying risk thresholds and committing to specific mitigations at each of those thresholds and being prepared to pause development or deployment if those mitigations are not in place. Effective risk management in frontier AI will require processes across a range of risk identification and mitigation measures, including six that many leading companies committed to in July 2023:
    \item Model Evaluations and Red Teaming can help assess the risks AI models pose and inform better decisions about training, securing, and deploying them. As new capabilities and risks can appear as frontier AI models are developed and deployed, evaluating models for several sources of risk and potential harmful impacts throughout the AI lifecycle is vital. External evaluation by independent, third party evaluators can also help to verify claims around the safety of frontier AI systems.
    \item Model Reporting and Information Sharing increases government visibility into frontier AI development and deployment. Information sharing also enables users to make well informed choices about whether and how to use AI systems. Practices involve sharing different information about their internal processes, safety and security incidents, and specific AI systems with different parties, including governments, other frontier AI organisations, independent third parties, and the public, as appropriate.
    \item Security Controls Including Securing Model Weights are key underpinnings for the safety of an AI system. If they are not developed and deployed securely, AI models risk being stolen or leaking secret or sensitive data, potentially before important safeguards have been applied. It is important to consider the cyber security of AI systems, as well as models in isolation, and to implement cyber security processes across their AI systems, including their underlying infrastructure and supply chains.
    \item Reporting Structure for Vulnerabilities enables outsiders to identify safety and security issues in an AI system. This is analogous to how organisations often set up ‘bug bounty programs’ for vulnerabilities in software and IT infrastructure. Practices include setting up a vulnerability management process that would cover many vulnerabilities - such as jailbreaking and prompt injection attacks - and have clear, user-friendly processes for receiving vulnerability reports.
    \item Identifiers of AI-generated Material provide additional information about whether content has been AI generated or modified. This can help prevent the creation and distribution of deceptive AI-generated content. Investing in developing techniques to identify AI-generated content is important in a highly nascent field, as well as exploring the use of approaches such as watermarks and AI output databases.
    \item Prioritising Research on Risks Posed by AI will help identify and address the emerging risks posed by frontier AI. Frontier AI organisations have a particular responsibility and capability to conduct AI safety research, to share their findings widely, and to invest in developing tools to address these risks. Collaboration with external researchers, independent research organisations, and third-party data owners will also be key to assessing the potential downstream social impacts of their systems. Risk management will likely require further measures than those already committed to, however. We suggest two additional processes and associated practices:
    \item Preventing and Monitoring Model Misuse is important as, once deployed, AI systems can be intentionally misused for harmful outcomes. Practices include establishing processes to identify and monitor model misuse, as well as implementing a range of preventative measures, and continually reviewing their effectiveness and desirability over time. Given the serious risks that misuse of frontier AI may pose, preparations could also be made to respond to potential worst-case misuse scenarios.
    \item Data Input Controls and Audits can help identify and remove training data likely to increase the dangerous capabilities their frontier AI systems possess, and the risks they pose. Implementation of responsible data collection and cleaning practices can help to improve the quality of training data before it is collected. Careful audits of training data both by frontier AI organisations themselves and external actors can also enable the identification of potentially harmful or undesirable data in training datasets. This can inform subsequent mitigatory actions, such as the removal of this data.
    \item Scan for novel or emerging risks – Proactively identify and address potential novel or emerging risks from foundation/ frontier models.
    \item Practise responsible iteration – Practice responsible iteration to mitigate potential risks when developing and deploying foundation/frontier models, through both internal testing and limited external releases.
    \item Assess upstream security vulnerabilities – Identify and address potential security vulnerabilities in foundation/frontier models to prevent unauthorised access or leaks.
    \item Produce a “Pre-Systems Card” – Disclose planned testing, evaluation, and risk management procedures for foundation/frontier models prior to development.
    \item Establish risk management and responsible AI structures for foundation models. – Establish risk management oversight processes and continuously adapt to address real world impacts from foundation/frontier models.
    \item Internally evaluate models for safety – Perform internal evaluations of models prior to release to assess and mitigate for potential societal risks, malicious uses, and other identified risks.
    \item Conduct external model evaluations to assess safety – Complement internal testing through model access to third-party researchers to assess and mitigate potential societal risks, malicious uses, and other identified risks.
    \item Undertake red-teaming and share findings – Implement red teaming that probes foundation/frontier models for potential malicious uses, societal risks and other identified risks prior to release. Address risks and responsibly disclose findings to advance collective knowledge.
    \item Publicly report model impacts and "key ingredient list" – Provide public transparency into foundation/frontier models’ “key ingredients” testing evaluations, limitations and potential risks to enable cross-stakeholder exploration of societal risks and malicious uses.
    \item Provide downstream use documentation – Equip downstream developers with comprehensive documentation and guidance needed to build safe, ethical, and responsible applications using foundation/frontier models. (Note: It is well understood downstream developers play a crucial role in anticipating deployment-specific risks and unintended consequences. This guidance aims to support developers in fulfilling that responsibility.)
    \item Establish safeguards to restrict unsafe uses – Implement necessary organisational, procedural and technical safeguards, guidelines and controls to restrict unsafe uses and mitigate risks from foundation/frontier models.
    \item Monitor deployed systems – Continuously monitor foundation/frontier models post-deployment to identify and address issues, misuse, and societal impacts.
    \item Implement incident reporting – Enable timely and responsible reporting of safety incidents to improve collective learning.
    \item Establish decommissioning policies – Responsibly retire foundation/frontier models from active use based on well defined criteria and processes.
    \item Develop transparency reporting standards – Collaboratively establish clear transparency reporting standards for disclosing foundation/frontier model usage and policy violations.
    \item Support third party inspection of models and training data – Support progress of third-party auditing capabilities for responsible foundation/frontier model development through collaboration, innovation and transparency.
    \item Responsibly source all labour including data enrichment – Responsibly source all forms of labour, including for data enrichment tasks like data annotation and human verification of model outputs.
    \item Conduct human rights due diligence – Implement comprehensive human rights due diligence methodologies to assess and address the impacts of foundation/frontier models.
    \item Enable feedback mechanisms across the AI value chain – Implement inclusive feedback loops across the AI value chain to ethically identify potential harms.
    \item Measure and disclose environmental impacts – Measure and disclose the environmental impacts resulting from developing and deploying foundation/frontier models.
    \item Disclose synthetic content – Adopt responsible practices for disclosing synthetic media and advance solutions for identifying other synthetic content
    \item Measure and disclose anticipated severe labour market risks – Measure and disclose potential severe labour market risks from deployment of foundation/frontier models.
    \item Monitoring and disrupting malicious state affiliated actors – investigate malicious actors in a variety of ways, including using our models to pursue leads, analyse how adversaries are interacting with our platform, and assess their broader intentions. Upon detection, OpenAI takes appropriate action to disrupt their activities, such as disabling their accounts, terminating services, or limiting access to resources.
    \item Working together with the AI ecosystem – OpenAI collaborates with industry partners and other stakeholders to regularly exchange information about malicious state-affiliated actors’ detected use of AI.
    \item Iterating on safety mitigations – Learning from real-world use (and misuse) is a key component of creating and releasing increasingly safe AI systems over time. We take lessons learned from these actors' abuse and use them to inform our iterative approach to safety.
    \item Public transparency – Highlight potential misuses of AI [link 1(opens in a new window), link 2] and share what we have learned about safety [link 1, link 2] with the industry and the public. As part of our ongoing efforts to advance responsible use of AI, OpenAI will continue to inform the public and stakeholders about the nature and extent of malicious state-affiliated actors’ use of AI detected within our systems and the measures taken against them, when warranted.
    \item Data Curation Practices – Prior to all training stages, we take various steps to mitigate potential downstream harms through data curation and careful data collection. We filter training data for high-risk content and to ensure training data is sufficiently high quality.
    \item Harm-inducing queries – To ensure broad coverage of harm-inducing queries, we enumerate approximately 20 harm types (e.g. hate speech, providing ungrounded medical advice, suggesting dangerous behaviour) across a wide variety of use cases.
    \item Supervised fine-tuning – Given the above harm-inducing queries, we create SFT data to demonstrate the safe and helpful responses for these queries. This includes human collections as well as a custom data generation recipe loosely inspired from Constitutional AI (Bai et al., 2022b), where we inject variants of Google’s content policy language as “constitutions”, and utilise language model’s strong zero-shot reasoning abilities (Kojima et al., 2022) to revise responses and choose between multiple response candidates.
    \item Reinforcement learning during human feedback – We also applied RLHF for the harm inducing queries, where we curated queries and model responses based on both observed loss patterns and our overall safety policy taxonomy, and then collected safety-specific preference data to be included into the overall RL reward model training mixture.
    \item Model-level Red Teaming – We apply state-of-the-art red teaming, a form of adversarial testing where adversaries launch an attack on an AI system, in order to test post-trained Gemini models for a range of vulnerabilities (e.g., cybersecurity) and social harms as defined in the safety policies. Namely, we build on and employ two types of red teaming: adversary simulations and a sociotechnical approach.
    \item Adversary simulations (unstructured testing) – are designed to emulate real-world adversaries and their approach to attacking models and associated systems, focusing on security, safety, and privacy failures. We combined in-house expertise with external experts to explore classes of vulnerabilities
    \item Structured Red Teaming – our second type of red teaming technique of Gemini models, takes a sociotechnical approach and makes three changes compared to SOTA red teaming techniques. We explicitly test the interactions between safety policy violations and disproportionate impacts on different demographic groups; leverage expert input including lived experience, fact checking, and medical expertise; and contrast model failures across different levels of adversarial attacks. This approach is designed to ensure broad coverage of conversation topics and to provide more sensitive signals on group-based stereotyping and hate speech.
    \item Reduce the prevalence of certain kinds of content that violate our usage policies (such as inappropriate erotic content) in our pre-training dataset
    \item Fine-tune the model to refuse certain instructions such as direct requests for illicit advice
    \item Leverage data from prior model usage to reduce the surface area of adversarial prompting or exploits
    \item Train a range of classifiers on new risk vectors and incorporate these into monitoring workflow to enforce API usage policies
\end{enumerate}

\end{document}